\documentclass[journal,a4paper]{IEEEtran}
\usepackage{amsmath,amssymb,amsfonts}
\usepackage{algorithmic}
\usepackage{algorithm}
\usepackage{array}
\usepackage[caption=false,font=normalsize,labelfont=sf,textfont=sf]{subfig}
\usepackage{textcomp}
\usepackage{stfloats}
\usepackage{url}
\usepackage{verbatim}
\usepackage{graphicx}
\usepackage{cite}
\usepackage{xcolor}
\usepackage{pifont}
\usepackage{cases}
\usepackage{multirow}
\usepackage{eqparbox}
\usepackage{geometry}
\begin{document}

\title{Secured and Cooperative Publish/Subscribe Scheme in Autonomous Vehicular Networks}


\author{Yuntao~Wang, Zhou~Su, \IEEEmembership{Senior Member, IEEE}, Qichao~Xu, Tom~H.~Luan, \IEEEmembership{Senior Member, IEEE}, and Rongxing~Lu, \IEEEmembership{Fellow, IEEE}
\thanks{Yuntao Wang, Zhou Su, and Tom H. Luan are with the School of Cyber Science and Engineering, Xi'an Jiaotong University, Xi'an, China}
\thanks{Qichao Xu is with the School of Mechatronic Engineering and Automation, Shanghai University, Shanghai, China}
\thanks{Rongxing~Lu is with the Faculty of Computer Science, University of New Brunswick, Fredericton, Canada}
}

\markboth{Journal of \LaTeX\ Class Files,~Vol.~XX, No.~X, XX~2021}%
{Wang \MakeLowercase{\textit{et al.}}: Secured and Cooperative Publish/Subscribe Scheme in Autonomous Vehicular Networks}


\maketitle
\begin{abstract}
In order to save computing power yet enhance safety, there is a strong intention for autonomous vehicles (AVs) in future to drive collaboratively by sharing sensory data and computing results among neighbors. However, the intense collaborative computing and data transmissions among unknown others will inevitably introduce severe security concerns.
Aiming at addressing security concerns in future AVs, in this paper, we develop SPAD, a secured framework to forbid free-riders and {promote trustworthy data dissemination} in collaborative autonomous driving. 
Specifically, we first introduce a publish/subscribe framework for inter-vehicle data transmissions{. To defend against free-riding attacks,} we formulate the interactions between publisher AVs and subscriber AVs as a vehicular publish/subscribe game, {and incentivize AVs to deliver high-quality data by analyzing the Stackelberg equilibrium of the game. We also design a reputation evaluation mechanism in the game} to identify malicious AVs {in disseminating fake information}.
{Furthermore, for} lack of sufficient knowledge on parameters of {the} network model and user cost model {in dynamic game scenarios}, a two-tier reinforcement learning based algorithm with hotbooting is developed to obtain the optimal {strategies of subscriber AVs and publisher AVs with free-rider prevention}.
Extensive simulations are conducted, and the results validate that our SPAD can effectively {prevent free-riders and enhance the dependability of disseminated contents,} compared with conventional schemes. 
\end{abstract}


\IEEEpeerreviewmaketitle
\section{Introduction}\label{sec:INTRODUCTION}
\IEEEPARstart{W}{ith} the advance of self-driving technologies, autonomous vehicles (AVs) are shifted from a fiction to an exciting practice with the promise to build future transportation systems with much-reduced traffic jams, much-improved road safety, and more intelligent vehicular services.
As reported in \cite{FutureAVs}, AVs are projected to reduce 90 percent of traffic deaths which are {due} to human errors and saving 30,000 lives a year.
Nevertheless, the success of AVs heavily relies on a multitude of onboard sensors to perceive surroundings and make real-time driving decisions.
Due to the intrinsic limitations of onboard sensors and processing capacity of a single AV, the road to fully autonomous driving is still fraught with challenges.

\begin{figure}[!t]
\centering
  \includegraphics[width=8.8cm]{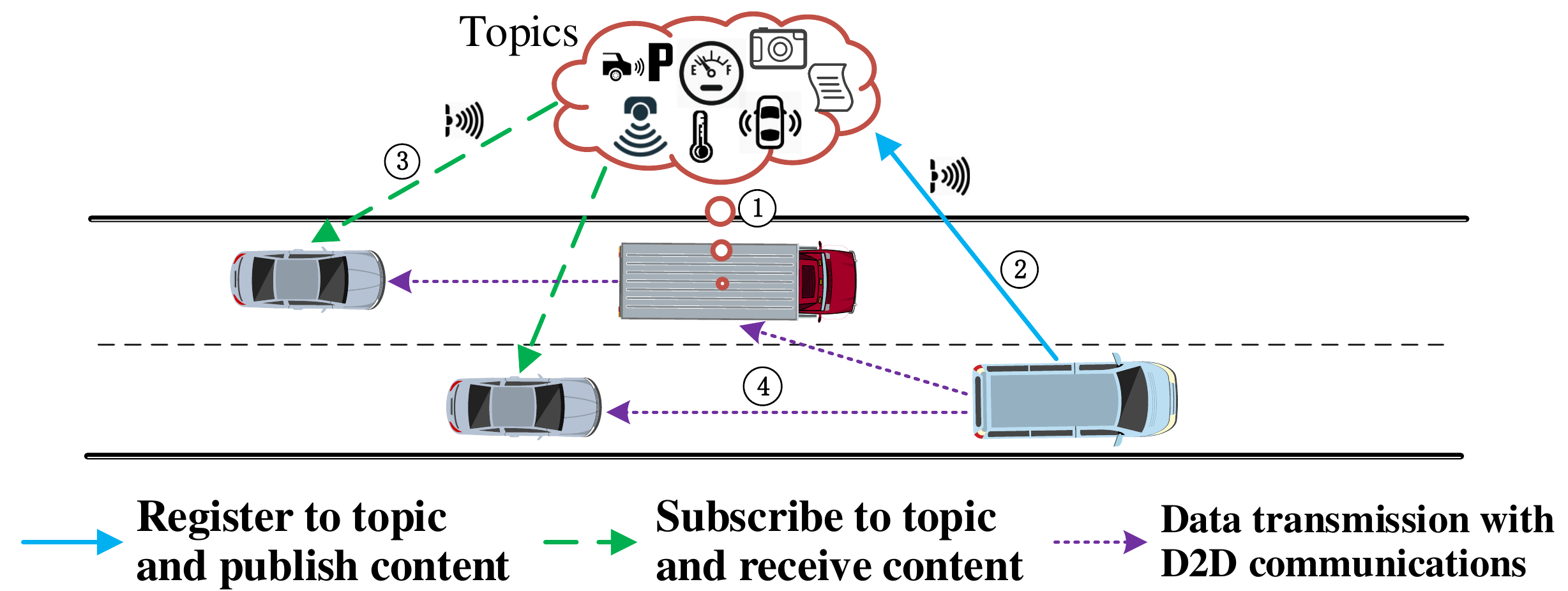}\\
  \caption{An example scenario of pub/sub framework in collaborative autonomous driving. (\ding{172}: {All contents to be shared within a fleet are tagged with topic strings and classified into various topics which are maintained by the master vehicle (e.g., topic creation and deletion, and notification of new contents to all subscribers); \ding{173}: A vehicle registers itself as a data source in certain topics and publishes individual sensory data (e.g., 3D point cloud data captured by onboard lidars) or processed results of the sensory information with tags to the registered topic;} \ding{174}: Neighboring vehicles {use string matching methods to search their interested topics and send subscriptions to} the master vehicle; \ding{175}: Subscribers {continually} retrieve {their} subscribed contents from the {corresponding} publisher using inter-vehicular {device-to-device (D2D)} communications.)} \label{fig:introduction}
\end{figure}

Cooperative autonomous driving, which is essentially to share driving statuses, sensory data, and computing results among neighboring {collaborators and road-side infrastructures}, has become crucial to improve driving accuracy and safety of single vehicles in the complicated and fast-changing driving environments \cite{Loke2019coop,8967210,Sridhar2019coop}.
For instance, cooperative AVs (CAVs) can construct a leader-follower formation, where a leader performs as the ``eye" of the platoon and shares its sight with the remaining CAVs (i.e., followers), and the followers can extend their awareness range to blind spots or areas beyond onboard capabilities with saved computing power \cite{Soni2018formation}.
{To enhance the communication efficiency under cooperative autonomous driving, the publish/subscribe (pub/sub) paradigm is utilized to deliver published information only to CAVs whose subscribed interests correspond to it \cite{9288629}.} The pub/sub paradigm has been implemented as the fundamental communication pattern in robot operating system as well as that in AVs \cite{liu2023ros,Sridhar2019coop,Kampmann2019ros}, e.g., Baidu Apollo operating system \cite{Apollo}. {By building a flexible asynchronous communication protocol, the pub/sub paradigm can efficiently disseminate information among CAVs with highly dynamic interactions \cite{6676844,Li2019fog}. For instance, by exploiting the relatively stable vehicle formation in a fleet, a CAV can continuously perceive its driving surroundings by subscribing the shared sight of certain neighbors.}
In a pub/sub framework as in Fig. \ref{fig:introduction}, a vehicle who is to share (publish) the sensory data or computing results would register {at certain topics} as a data source. {Neighboring} vehicles who are interested in the information need to subscribe to the topic so as to continually receive the published information from the publisher. As a result, data are retrieved in an on-demand approach; each vehicle can publish certain topics of content (sensed data or {its processed outcomes}) and subscribe to other topics, leading to a mesh connected collaborative system.

However, the full deployment of pub/sub mechanism in autonomous driving still faces a series of challenges.
On one hand, resource-limited AVs may be selfish {and free-riding. They} will not contribute their content, or they will deliver meaningless and low-quality content if there is no sufficient validation or incentive.
On the other hand, malicious AVs may register as a publisher and publish dishonest or even harmful contents to mislead and interfere the normal driving activities of nearby AVs.
The absence of security validations may further reduce the willingness of AVs to participate in the cooperation.
Thus, it is still pressing to design a secure {data pub/sub scheme} for CAVs while promoting their cooperation in data contribution. 

Quite recently, a number of research works have been developed towards efficient and secure autonomous driving.
For example, in \cite{Jiang2020-1}, a novel authentication and key agreement mechanism is proposed to ensure secure remote control of AVs and guarantee secure data exchange between AVs and the cloud.
Besides, a mobile edge computing (MEC) based data trading scheme with debt-credit is presented in \cite{Liu2019-2}, where the competitive interactions between borrower vehicles and lender vehicles are formulated as a Stackelberg game to maximize the profit of both sides.
Meanwhile, in \cite{Chen2019-3}, a coalitional game based framework is devised to optimize the cooperative behaviors of neighboring vehicles and facilitate the popular content distribution.
However, most of the current works cannot be directly applied for {cooperative autonomous driving} with the following aspects.
{First,} AVs in a fleet need to continuously receive the subscribed information from neighbors until the formation changes, while the implementation of pub/sub paradigm in AV fleets is seldom considered.
{Besides,} AVs are typically equipped with diverse types of sensors, whereas the diversified sensing and processing capabilities {and costs} are not sufficiently taken into account in most of the existing works.
{Second, AVs commonly exhibit diverse behaviors in the pub/sub process. For example, free-riding AVs may refuse to pay and contribute data, and aggressive AVs may disseminate forged or dishonest information. Few works consider the incentives for AVs' cooperative behaviors to encourage AVs' participation and high-quality data contribution.}
{Third, different from the assumption in current game-based incentive approaches, it is not readily available for AVs to accurately obtain parameters of the dynamic network model and users' private cost model in time due to AVs' high mobility and privacy concerns. A learning-based incentive scheme to promote AVs' cooperation without the full knowledge of network and user parameters is needed.}
Therefore, it is still an open and vital issue to safeguard autonomous driving while motivating {AVs' cooperation} under the pub/sub framework.

In this paper, we develop SPAD, a {\underline{s}ecured \underline{p}ub/sub framework for collaborative \underline{a}utonomous \underline{d}riving}. 
In specific, we first apply the pub/sub mechanism for inter-vehicular communications and MEC for vehicle-to-infrastructure (V2I) communications.
After that, {to stimulate CAVs' participation and high-quality content contribution,} a vehicular pub/sub game is formulated {by modeling} the interactions of CAVs in {publishing/subscribing}. In the devised game model, each subscriber CAV is the game-leader to decide the payment strategy of subscribed content and the publisher CAV is the game-follower to make the strategy on quality of published content. {By analyzing the Stackelberg equilibrium (SE) of the game, competitive subscriber AVs can determine their optimal payment strategies to encourage high-quality content services of publisher CAVs, thereby preventing free-riding threats.
Moreover, based on vehicle behaviors and social roles,} a reputation mechanism is developed in the game model to identify malicious CAVs {which publish dishonest information.} 
Besides, for the lack of knowledge of accurate network parameters in the highly dynamic environment, the interactions between publisher CAVs and subscriber CAVs are formulated as finite Markov decision processes (MDPs). In the dynamic game, a two-tier policy hill-climbing (PHC) based reinforcement learning algorithm is devised to derive the optimal policies of CAVs with accelerated convergence rate via trials. The hotbooting method is also exploited in PHC to initialize the Q-tables with experiences in similar scenarios to avoid random exploration and improve learning efficiency.
The main contributions of this work are threefold as follows:
\begin{itemize}
\item \emph{System:} We present SPAD, a practical MEC-assisted pub/sub framework in cooperative autonomous driving for {secure} data {publishing/subscribing} for CAVs with untrusted neighbors. We formulate the interactions between the publisher CAV and subscriber CAVs as a vehicular pub/sub game and derive the optimal {strategy} for each player with maximized individual utility in both static and dynamic games.
\item \emph{Scheme:} According to CAVs' social roles and vehicle behaviors, we evaluate CAVs' trustworthiness by reputation mechanism to defend against {false data publishing attacks}. A reputation increase can be regarded as a reward to motivate CAVs to behave legitimately.
    Furthermore, {the SE of the static vehicular pub/sub game and its stability are} analyzed, where each publisher CAV is motivated to offer high-quality content and the selfishness of CAVs is suppressed.
    A two-tier hotbooting PHC algorithm is also developed in the dynamic game to acquire the optimal payment strategies of subscriber CAVs and the optimal quality strategies on content services of publisher CAVs with improved learning efficiency.
\item \emph{Validation:} We evaluate the effectiveness of SPAD through extensive simulations. It is demonstrated that our SPAD can attain higher quality of {published data}, enhanced {dependability of subscribed data} for CAVs, improved user utilities, and faster convergence rate, by comparing with other existing schemes.
\end{itemize}

The remainder of this paper is organized as follows. Section~\ref{sec:RELATEDWORK} outlines the related work. Section~\ref{sec:SYSTEMMODEL} introduces the system model.
We formulate the vehicular pub/sub game in Section~\ref{sec:SPAD}, and analyze the SE of the static game in Section~\ref{sec:game}.
In Section~\ref{sec:learning}, the reinforcement learning-based optimal strategy decision {to solve} the dynamic game is elaborated.
Performance evaluation is given in Section~\ref{sec:SIMULATION}. Section~\ref{sec:CONSLUSION} concludes this paper and points out the future work.

\section{Related Works}\label{sec:RELATEDWORK}

\subsection{Incentive of Autonomous Driving}\label{subsec:relatedwork1}
Incentive mechanism is fundamental in a collaborative system in general and collaborative autonomous driving in specific.
Fabiani \emph{et al}. \cite{2-2} study the automated driving coordination problem for multiple selfish AVs on multi-lane highways through generalized mixed-integer potential game approaches. 
Su \emph{et al}. \cite{Su2018} formulate a market-based mechanism to optimally incentivize AVs to contribute individual computing resources for autonomous driving decision making.
Two dynamic game paradigms including a Stackelberg game and a zero-sum game are exploited by Ji \emph{et al}. \cite{2-4} to improve the stability and robustness for path tracking of connected AVs.
Tian \emph{et al}. \cite{8651307} formulate an evolutionary game based framework for channel access optimization in cognitive radio empowered vehicular networks, where a delayed pricing mechanism with discretized replicator dynamics is designed to improve the evolutionary stability and efficiency of Nash equilibrium.
{However, existing works} mainly focus on a generic mesh connected vehicular topology without considering the detailed underlying inter-vehicular communication scheme. Our work investigates on the pub/sub mechanism which is widely implemented in real-world commercial deployments such as Baidu Apollo project \cite{Apollo} and is more practical in autonomous driving.

{Besides,} reinforcement learning, as a branch of machine learning technique, has been widely applied to offer incentives in vehicular networks.
To improve driving efficiency and safety at signalized intersections, Zhou \emph{et al}. \cite{8848852} propose an intelligent car following mechanism based on reinforcement learning for connected AVs to schedule driving behaviors in real-time by exploiting the shared information of neighbors.
Zhao \emph{et al}. \cite{9173810} utilize deep reinforcement learning techniques to acquire the optimal long-term sensing strategy of vehicles under sensing budget in dynamic vehicular social networks.
Q-learning is a typical model-free reinforcement learning algorithm.
By considering future network states in the learning phase, Zhou \emph{et al}. \cite{8944281} devise a Q-learning based algorithm for efficient resource allocation and adaptive time division duplex configuration in 5G vehicular networks.
{However, few works consider the use of reinforcement learning techniques to defend against free-riders in incentive mechanism design under highly dynamic autonomous vehicular networks for enhanced security of publication/subscription services.}

\subsection{Cyber Security for Autonomous Driving}\label{subsec:relatedwork3}
Recent works on cyber security of autonomous driving mainly focus on cryptographic mechanisms.
Parkinson \emph{et al}. \cite{Park2017} review the applications of cryptographic systems to guarantee information security and privacy for connected AVs.
Through cryptographic protocols, Karnouskos \emph{et al}. \cite{Karn2018} study data integrity and privacy issues for real-world deployment of hyper-connected AVs with vehicle-to-everything (V2X) support. 
Lai \emph{et al}. \cite{Lai2017} present a secure cooperative content downloading scheme in highways by rewarding proxy neighboring vehicles to collect data fragments to obtain the complete data via asymmetric encryption and integrity verification. 
{Nonetheless, cryptographic mechanisms} cannot cope with the threats arisen from inside attackers well (i.e., the attacks targeting at the inside nodes). For example, legitimate AVs with malfunctioning sensors may send incorrect data such as fake warnings to neighbors. Moreover, legitimate AVs may be compromised to disseminate false information.

Reputation and trust model is another efficient tool to safeguard vehicular networks. {Existing reputation models can be mainly divided into two kinds: \emph{data-centric} and \emph{entity-centric}. In data-centric reputation models, it concentrates on computing the trustworthiness of data according to the context of events (or behaviors), event types, reports of the same event, etc. In the literature, the Bayesian inference model has been widely adopted to build data-centric reputation models in vehicular networks \cite{Maga2019,8358773}, where the probability distribution of binary events (e.g., bad or good) is assumed to follow the beta distribution. 
Entity-centric approaches focus on evaluating the trustworthiness of entities (e.g., vehicles) based on multifaceted methods \cite{5641621} (e.g., local experience, priority, and role), recommendation-based methods \cite{9632356} (e.g., votes given from vehicles and infrastructures), etc.
By contrast, we build a hybrid reputation model which considers both the trustworthiness of vehicles and the data they exchange, by dynamically updating vehicles' trust to evaluate the trustworthiness of their delivered data. Thereby, the phenomenon that trustworthy vehicles may deliver false data in the presence of adversaries can be mitigated. Besides, we further design an improved Bayesian inference model to improve the accuracy and robustness of our proposed reputation mechanism.}

{Other works also exploit blockchain and physical control methods to safeguard cooperative autonomous driving. Xing \emph{et al}. \cite{9288629} propose a secure pub/sub scheme with the assistance of truck platoons in autonomous vehicular networks, where truck platoons act as distributed brokers (i.e., master vehicles) of the pub/sub system. Besides, they deploy a vehicular blockchain network to resist deception, denial-to-pay, and denial-to-forward-content misbehaviors by offing decentralized, transparent, and immutable ledgers. By using an adaptive control method, Petrillo \emph{et al}. \cite{8967210} develop a secure and resilient leader tracking strategy for a homogeneous AV platoon to mitigate cyber threats including spoofing, message falsification, denial-of-service (DoS), and burst transmission. The effectiveness of the proposed strategy in \cite{8967210} is analytically proved by the Lyapunov-Krasovskii approach under reasonable assumptions. Different from the above works, we mainly focus on the defense of free-riding misbehaviors of AVs, which can impede the practical deployment of cooperative autonomous driving and is absent in most of the existing works.}

In the light of existing works, our work studies the secure data transmissions in cooperative autonomous driving by exploiting the feature of CAV networks, i.e., distributed pub/sub and MEC communications. In addition, a reputation mechanism and learning-based incentive mechanism are investigated to stimulate {trusted and high-quality data contribution of AVs with better adaptation to the fast-changing network environment}. 

\begin{table}[!t]
\caption{Summary of Notations}\label{table0}\centering
\vspace{-2mm}
\resizebox{1.02\linewidth}{!}{
\begin{tabular}{|c|l|}
\hline 
\textbf{Notation} & \textbf{Description} \\   \hline
    $\mathcal{K}$&Set of CAV fleets. \\
    $\mathcal{N}_k$&Set of CAVs in fleet $k \in \mathcal{K}$. \\
    $\mathcal{M}$&Set of MEC nodes. \\
    $\mathcal{T}$&Set of time slots. \\
    $\mathcal{O}_k$&Set of topics in fleet $k$ within time horizon $\mathcal{T}$. \\
    $\mathcal{G}$&Set of sensors mounted on CAVs. \\
    $h_k$ &Master CAV (i.e., broker of pub/sub system) in fleet $k$. \\
    $\mathcal{I}_k$&Set of publisher CAVs in fleet $k$. \\
    $\mathcal{J}_k$&Set of subscriber CAVs in fleet $k$. \\
    \multirow{2}*{$\mathcal{C}_i$}
        &Set of published contents of publisher CAV $i \in \mathcal{I}_k$ \\ ~ & \qquad at time slot $t \in \mathcal{T}$.   \\
    $\mathcal{J}_{c}$&Subscriber CAV group of a specific content $c$. \\
    $sc_{n,g}$ &Sensing capacity of CAV $n\!\in\! \mathcal{N}$ on sensor of type $g\!\in\! \mathcal{G}$. \\
    $pc_{n}$ &Processing capability of CAV $n$. \\
    ${raw{_c}}$&The raw sensory part of content $c$. \\
    ${{result_{c}}}$&The processed outcome part of content $c$. \\
    \multirow{2}*{$\beta_{j,c}$}
        &Binary preference variable of subscriber CAV $j$ in \\ ~ & \qquad subscribing ${raw{_c}}$ or ${{result_{c}}}$ of content $c$. \\
    ${Q_{i,c}^1},{Q_{i,c}^2}$ &Quality of sensory data $raw_c$\big/processed result $result_c$. \\
    \multirow{2}*{${\mathbf{q}_{i,c}}$}
        &Quality of content service (QoCS) vector of publisher \\ ~ & \qquad CAV $i$ in sharing content $c$.   \\
    \multirow{2}*{${\mathbf{p}_{i,c}}$}
        &Payment vector of subscriber CAV group $\mathcal{J}_c$ for attaining \\ ~ & \qquad content $c$.   \\
    ${f_c}$ &Popularity degree of content $c \in \mathcal{C}_k$. \\
    \multirow{2}*{$MulticastAdd_c$}
        &Multicast group address assigned by master CAV $h_k$ to \\ ~ & \qquad multicast content $c$ to all subscriber CAVs in $\mathcal{J}_c$. \\
    $R_n$&Reputation value of CAV $n$. \\
    ${\alpha _{i,c}}$ &Satisfaction coefficient of subscriber CAVs of content $c$. \\
    \multirow{2}*{$\varepsilon _{i,\pi_c}^1$}
        &Cost parameter of publisher CAV $i$ with the highest QoCS \\ ~ & \qquad in contributing $raw_c$ using type-$\pi_c$ sensor. \\
    \multirow{2}*{$\varepsilon _i^2$}
        &Cost parameter of publisher CAV $i$ with the highest QoCS \\ ~ & \qquad in computing $result_c$. \\
    ${\mathcal{U}_i}\left( {{{\bf{q}}_i},{{\bf{p}}_i}} \right)$&Utility function of publisher CAV $i$ in content sharing. \\
    ${\mathcal{U}_{\mathcal{J}_c}}\left({\mathbf{q}_{i,c}},{\mathbf{p}_{i,c}}\right)$&Utility function of subscriber CAV group $\mathcal{J}_c$. \\
    ${\bf{z}}_{i,c}^t$&QoCS state of publisher CAV $i$ on content $c$ at time slot $t$. \\
    $\mathbb{{Q}}({\bf{z}}_{i,c}^t, {\bf{p}}_{i,c}^t)$&Q-function of subscriber CAV group $\mathcal{J}_{c}$. \vspace{0.2mm}\\
    $\pi ( \mathbf{z}_{i,c}^{t},\mathbf{p}_{i,c}^{t} )$&Mixed-strategy table of subscriber CAV group $\mathcal{J}_{c}$. \vspace{0.2mm}\\
    $\tilde{\bf{z}}_{i,c}^t$&Payment state of subscriber CAV group $\mathcal{J}_{c}$ at time slot $t$. \\
    $\tilde{\mathbb{{Q}}}(\tilde{\bf{z}}_{i,c}^t, {\bf{q}}_{i,c}^t)$&Q-function of publisher CAV $i$. \vspace{0.2mm}\\
    $\tilde{\pi}(\tilde{\bf{z}}_{i,c}^t, {\bf{q}}_{i,c}^t)$&Mixed-strategy table of publisher CAV $i$. \vspace{0.3mm}\\
\hline 
\end{tabular} }
\end{table}
\section{System Model}\label{sec:SYSTEMMODEL}
In this section, we introduce the system model by discussing on the network model, mobility model, content model, pub/sub model, and security model, respectively. A summary of notations used in the remaining of this paper is presented in Table~\ref{table0}.

\subsection{Network Model}\label{subsec:networkmodel}
\begin{figure}[!t]\setlength{\abovecaptionskip}{-0.05cm}\vspace{-3mm}
\centering
  \includegraphics[height=5.8cm]{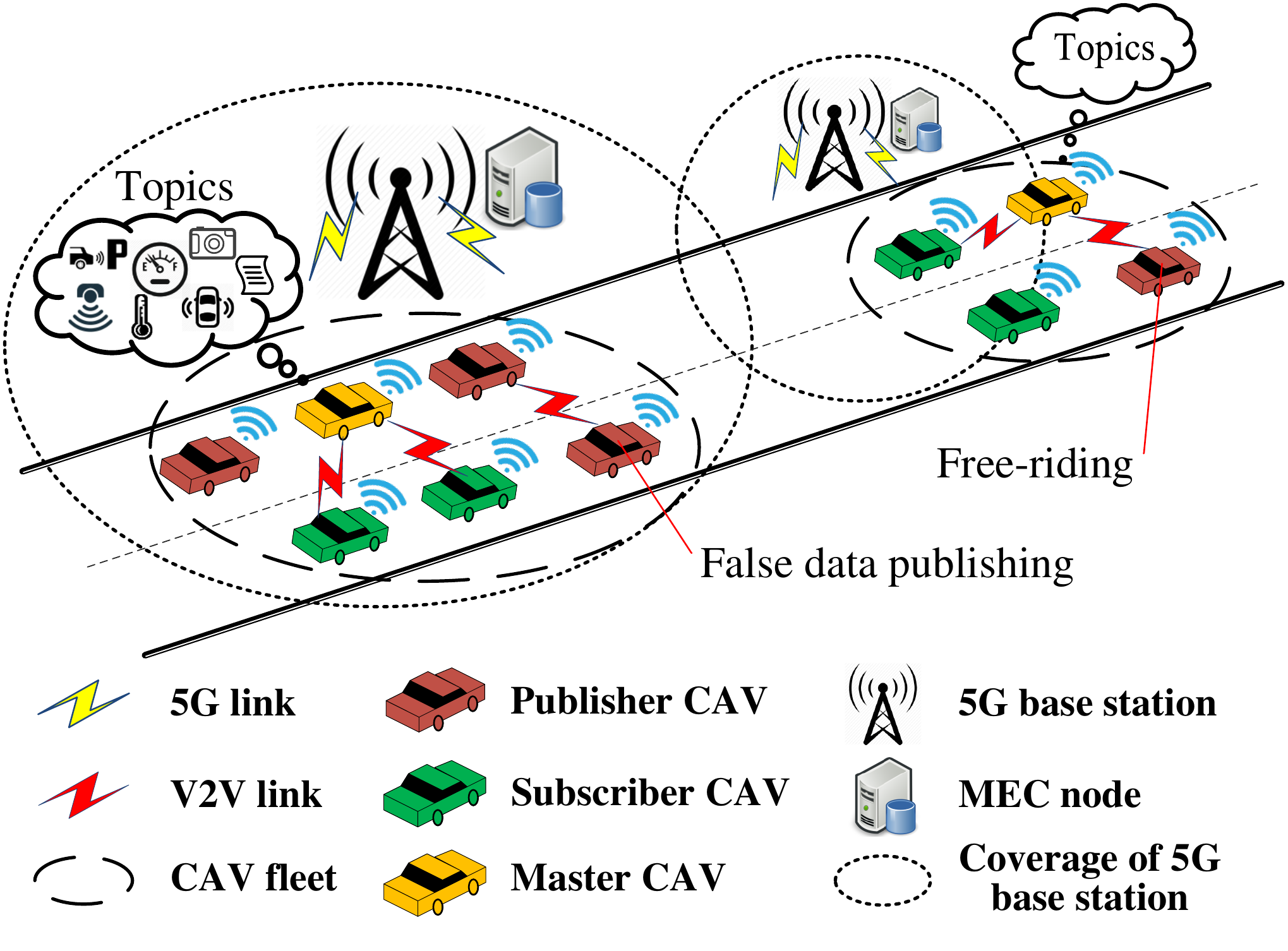}\\ 
  \caption{System model of SPAD.}\label{fig:systemmodel}\vspace{-4mm}
\end{figure}

Fig.~\ref{fig:systemmodel} shows the scenario of cooperative autonomous driving considered, which includes fleets of neighboring CAVs, MEC nodes, and 5G base stations.

\emph{\textbf{CAVs}.} In a given investigated area, each CAV in the set $\mathcal{N} = \{1,\cdots,n,\cdots,N\}$ is equipped with a 5G-V2X integrated onboard unit device (OBU) to support both vehicle-to-vehicle (V2V) communications and V2I communications \cite{Karn2018}.
For improved safety, driving accuracy, and fuel efficiency, a group of neighboring CAVs can form a CAV fleet\footnotemark[1] through vehicle formation controlled by self-driving programs (e.g., cooperative adaptive cruise control). 
\footnotetext[1]{In this work, we mainly focus on the security of data {publishing/subscribing} for generalized collaborative autonomous driving in the pub/sub framework, where the dynamic formation of CAV fleets can be seen in \cite{Loke2019coop,Soni2018formation,Zhai2019a-add}.}
Within a CAV fleet, CAVs can drive collaboratively on the road by exchanging sensory information with each other via OBUs. 
The set of CAV fleets is denoted by $\mathcal{K} = \{1,\cdots,k,\cdots,K\}$, and the set of CAVs in fleet $k$ is $\mathcal{N}_k = \{1,\cdots,n,\cdots,N_k\}$.
{According to works \cite{8967210,9288629}, we make a similar assumption that neighboring CAVs with the same destination and driving route can drive cooperatively and form a CAV fleet. Thereby,} the vehicle formation and topology within each CAV fleet is relatively stable in a certain {time periods}. 
There exist the following three kinds of CAVs in the pub/sub system within a CAV fleet:
\begin{itemize}
  \item \emph{\textbf{Master CAVs}.} As the broker of pub/sub system\footnotemark[2], each master CAV $h_k$ maintains all topics within fleet $k \in \mathcal{K}$. For all contents to be published in fleet $k$, master CAV $h_k$ generates the metadata for each of them and classifies them into different topics. Let $\mathcal{O}_k = \{1,\cdots,o,\cdots,O_k\}$ and $\mathcal{C}_k = \{1,\cdots,c,\cdots,C_k\}$ denote the sets of topics and published contents in CAV fleet $k$ within time slots $\mathcal{T} = \{1,\cdots,t,\cdots,T\}$, respectively. 
      Each master CAV $h_k$ maintains {the publish/subscription information} for every topic $o \in \mathcal{O}_k$, notifies all subscribers about the new content published in their subscribed topics, and assists to establish V2V connections between publishers and subscribers for content delivery. {The buffering capacity of master CAV $k$ is denoted by $B_k$. Note that in the fast-changing driving environment, subscriber CAVs intend to receive the fresh published data instead of the old one. Therefore, the master CAV $k$ only needs to store the latest published data within a fixed time window $W_{o,k}$ for topic $o$ instead of storing all published data, where the value of $W_{o,k}$ depends on specific applications. Besides, master CAVs only store the metadata of published contents in topics to decrease communication and storage costs.}
      \footnotetext[2]{{Due to the high deployment cost of road-side infrastructures such as MEC nodes, not all roads are in the coverage of MEC nodes especially for highways and rural roads. Besides, due to the high mobility of CAVs, the dwell time of a CAV fleet in a MEC node can be very limited, resulting additional time costs in frequent data/service handover between MEC nodes. Therefore, we utilize the master CAV as the broker of pub/sub system.}}
  \item \emph{\textbf{Publisher CAVs}.} The set of publisher CAVs in fleet $k$ is defined as $\mathcal{I}_k = \{1,\cdots,i,\cdots,I_k\}$, where $\mathcal{I}_k \subseteq \mathcal{N}_k$. Each publisher CAV $i \in \mathcal{I}_k$ registers as a data source at master CAV $h_k$ and publishes individual contents to registered topics. The set of published contents of publisher CAV $i$ at time slot $t \in \mathcal{T}$ is denoted as $\mathcal{C}_i = \{1,\cdots,c,\cdots,C_{i}\}$. Let $D_i$ be the caching capacity of publisher CAV $i$, and $s_c$ be the data size of each published content $c$. Here, $\sum\nolimits_{c \in {\mathcal{C}_i}} {s_c \le {D_i}}$.
  \item \emph{\textbf{Subscriber CAVs}.} The set of subscriber CAVs in fleet $k$ is defined as $\mathcal{J}_k = \{1,\cdots,j,\cdots,J_k\}$, where $\mathcal{J}_k \subseteq \mathcal{N}_k$. Each subscriber CAV $j \in \mathcal{J}_k$ subscribes to its interested topics which are registered at master CAV $h_k$ to continually acquire desired contents, which is transmitted through V2V multicasting. To mitigate the cost of acquiring content $c \in {\mathcal{C}_i}$ contributed by publisher CAV $i$, subscriber CAVs of content $c$ can form the subscriber CAV group $\mathcal{J}_{c} = \{1,\cdots,j,\cdots,J_c\}$ {and evenly dividing the payment}. 
\end{itemize}

Each CAV $n \in \mathcal{N}$ is equipped with various types of sensors to perceive its surroundings, e.g., cameras, lidars, radars, etc. Assume that there exist $G$ types of sensors mounted on CAVs, the set of which is denoted as $\mathcal{G} = \{1,\cdots,g,\cdots,G\}$. Since CAVs have different sensing capacities on the same type of sensors, we set $sc_{n,g} \in [0,1]$ to indicate the sensing capacity of CAV $n$ on sensor of type $g$. Specifically, $sc_{n,g}=1$ means that CAV $n$ has the highest sensing capacity on sensor of type $g$, while $sc_{n,g}=0$ indicates that the sensing capacity of CAV $n$ on type-$g$ sensor is the lowest. Here, if CAV $i$ is not equipped with type-$g$ sensor, its sensing capacity on that type of sensor is zero, i.e., $sc_{n,g}=0,\forall n \in \mathcal{N},\forall g \in \mathcal{G}$.
Additionally, CAVs can have diversified processing capabilities in performing computation tasks. Let $pc_{n}\in [0,1]$ be the processing capability of CAV $n$. Here, $pc_{n}=1$ and $pc_{n}=0$ imply that CAV $n$ has the highest and lowest processing capability, respectively.

\emph{\textbf{MEC Nodes}.} The set of MEC nodes is denoted by $\mathcal{M} = \{1,\cdots,m,\cdots,M\}$. Each MEC node is deployed at a 5G base station to provide edge computing, edge caching, and V2I communication capacities for CAVs in its coverage to facilitate vehicular content services. The communication coverage of each MEC node $m \in \mathcal{M}$ is a circle with radius $radius_m$. Additionally, MEC node $m$ can serve as a publisher (e.g., publish on-road traffic events) or a subscriber in certain topics of CAV fleets in its coverage to facilitate inter-fleet data exchange. For instance, MEC node $m$ can receive security-critical contents by subscribing to related topics in a CAV fleet and {delivering} them to other CAV fleets that are in need of. 

\subsection{Mobility Model}\label{subsec:trafficmodel}
\begin{figure}[!t]\setlength{\abovecaptionskip}{-0.05cm}\vspace{-3mm}
\centering
  \includegraphics[height=3.6cm]{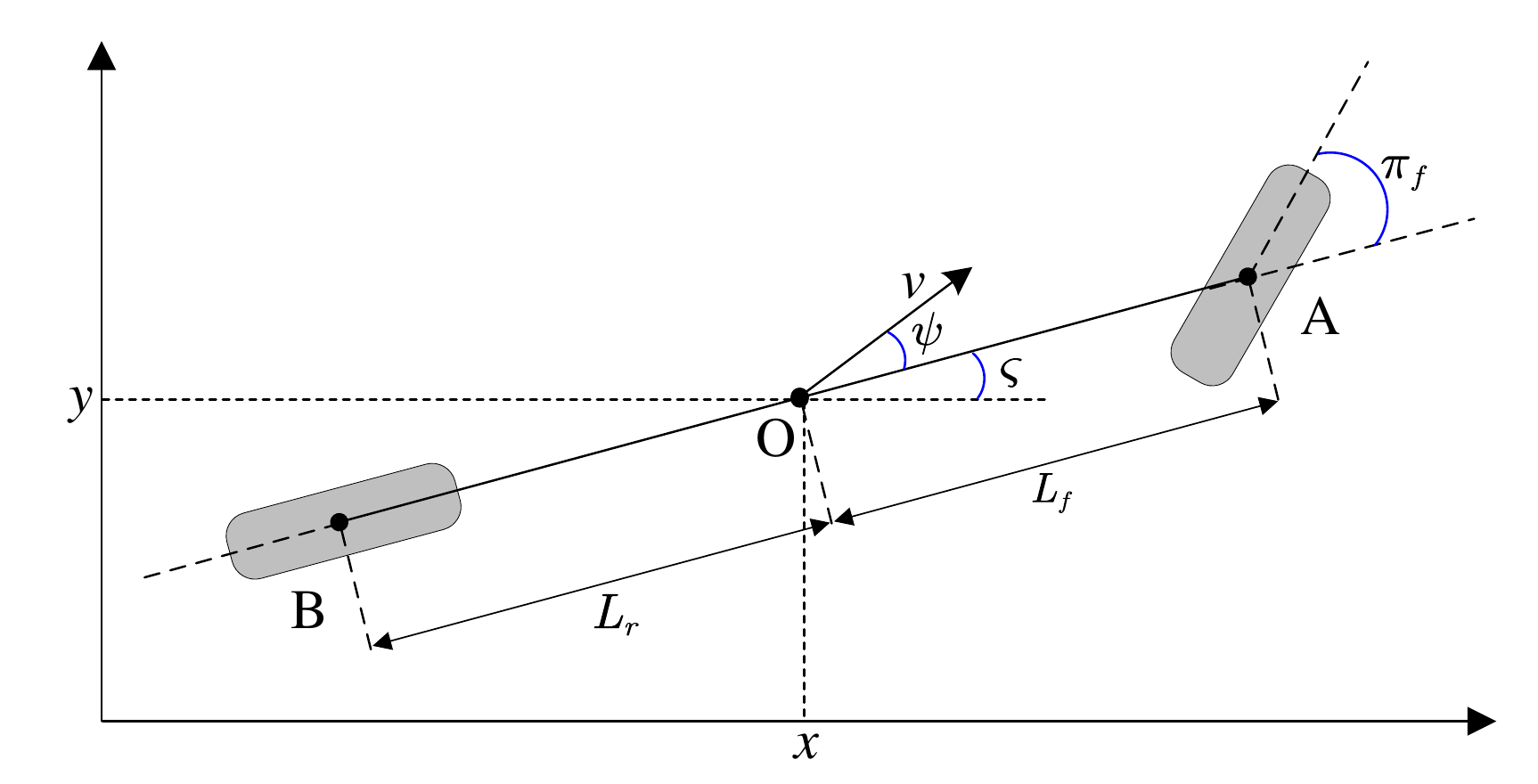}\\
  \caption{{Illustration of the kinematic bicycle model for CAVs. $A$, $B$, and $O$ are the center of front axle, center of rear axle, and center of gravity, respectively. $\varsigma$ is the heading angle. $\psi$ is the slip angle. $\pi_f$ is the front wheel steering angle.}}\label{fig:motionmodel}\vspace{-2mm}
\end{figure}

{Let ${t_s}$ and ${t_e} = {t_s} + \delta$ be the start and end time of time slot $t\in \mathcal{T}$, respectively. Since $\delta$ can be small enough, the instant status of CAV $n \in \mathcal{N}$ can be approximately fixed within each time slot $t$ but varies over different time slots, i.e., 
\begin{align}\label{status}
{{\bf{l}}_n}(t) = \left[ {{x_n}(t),{y_n}(t)},{v _n}(t),\varsigma_n(t) \right], \forall n \in \mathcal{N}, \forall t \in \mathcal{T},
\end{align}
where ${x_n}(t)$ and ${y_n}(t)$ are the horizontal coordinates of CAV $n$, ${v _n}(t)$ is the velocity of CAV $n$, and $\varsigma_n(t)$ is the inertial heading of CAV $n$. Once a CAV fleet $k$ is formed, all CAVs in this fleet will drive at the same velocity ${v _k^*}(t)$ to maintain desired formation shape and inter-vehicle safe distance $D_k^{*}$. For all CAVs in a fleet $k$, we have ${v _n}(t)={v _k^*}(t)$.
According to the kinematic bicycle model \cite{8429261}, as shown in Fig.~\ref{fig:motionmodel}, the motion dynamics of CAV $n$ can be characterized by the following equations:
\begin{align}\label{motion1}
{x_n}(t+1) = {x_n}(t) + {v _n}(t)\cos(\varsigma_n(t)+\psi_n(t))\delta,
\end{align}
\begin{align}\label{motion2}
{y_n}(t+1) = {y_n}(t) + {v _n}(t)\cos(\varsigma_n(t)+\psi_n(t))\delta,
\end{align}
\begin{align}\label{motion3}
{v _n}(t+1) = {v _n}(t) + a_n(t)\delta,
\end{align}
\begin{align}\label{motion4}
{\varsigma_n}(t+1) = {\varsigma_n}(t) + \frac{{v _n}(t)\sin(\psi_n(t))}{L_r}\delta.
\end{align}
Here, $a_n(t)$ is CAV $n$'s acceleration. $a_n(t)>0$ means stepping on the accelerator, $a_n(t)>0$ means stepping on the brake, and $a_n(t)=0$ means that the vehicle moves at the same speed. $\psi_n(t)$ is the slip angle of CAV $n$, which is determined by
\begin{align}\label{motion5}
{\psi_n}(t) = \tan^{-1}\left(\frac{L_r \tan(\pi_f(t))}{L_r + L_f} \right),
\end{align}
where $L_f$ and $L_r$ are the distances from front and rear axles to CAV's center of gravity, respectively. $\pi_f(t)$ is the front wheel steering angle. As the rear wheels cannot be steered in most AVs, we assume that only the front wheel can be steered \cite{8429261}. 
Based on Eqs. (\ref{motion1})--(\ref{motion5}), given the inputs (i.e., $a_n(t)$ and $\pi_f(t)$) at current time slot $t$, the status of any CAV in Eq. (\ref{status}) at the next time slot $t+1$ can be predicted for automatic vehicle control in a fleet.}

\subsection{Content Model}\label{subsec:contentmodel}
In CAV fleet $k$, each published content $c \in \mathcal{C}_k$ in topic $o \in \mathcal{O}_k$ has two parts, i.e., the raw sensory data ${raw{_c}}$ and the corresponding processed results ${{result_{c}}}$.
For instance, ${raw{_c}}$ can be the 3D point cloud data \cite{Zeng2018rt3d} captured by onboard lidars in real-time autonomous driving, while ${{result_{c}}}$ can be the relevant classification results through object recognition methods.
Hence, each content $c$ can be formulated as $c = \{{raw{_c}},\, {{result_{c}}} \}$.
Commonly, subscriber CAV who disbelieves the processed results ${{result_{c}}}$ computed by the publisher CAV, prefers to request the raw sensory data ${raw{_c}}$ and process it locally to improve driving safety.
Here, each subscriber CAV of content $c$ can either request the raw sensory data ${raw{_c}}$ or the processed results ${{result_{c}}}$ from the publisher CAV. Let $\beta_{j,c} = \{0,1\}$ denote the binary preference variable of subscriber CAV $j$ on content $c$.
If $\beta_{j,c}=1$, it indicates that subscriber CAV $j$ prefers requesting the raw sensory data of content $c$. Otherwise, $\beta_{j,c} = 0$ means that subscriber CAV $j$ prefers the computing results of content $c$.
We define $\mathcal{J}_c^1 = \{1,\cdots,j,\cdots,J_c^1\}$ and $\mathcal{J}_c^2 = \{1,\cdots,j,\cdots,J_c^2\}$ as the sets of subscriber CAVs of the raw sensing data $raw_c$ and the computing result $result_c$ of content $c$, respectively, where $\mathcal{J}_c^1 \cup \mathcal{J}_c^2 = \mathcal{J}_c$. 

Let $\mathbf{Q}_{i,c} = {\left[{Q_{i,c}^1}, {Q_{i,c}^2} \right]}$ be the quality vector of content $c$ published by CAV $i$, where ${Q_{i,c}^1}$ and ${Q_{i,c}^2}$ are the quality of sensory data $raw_c$ and computing result $result_c$, respectively. Intuitively, $\mathbf{Q}_{i,c}$ indicates the quality of CAV $i$'s sensory data or processing results. The meaning of content quality varies for different types of onboard sensors. For instance, it refers to the sensing quality (e.g., resolution, sharpness, {and contrast}) of camera sensors, and the computing accuracy (e.g., the accuracy of image classification) of in-built processors. Higher quality of delivered data can help subscribers obtain high-accurate information about their driving surroundings.
Here, $\mathbf{Q}_{i,c}$ is affected by the type of used sensor and the sensing/processing capacity of CAV $i$ in generating content $c$, namely,
\begin{align}
{{\bf{Q}}_{i,c}} = \left[{q_{i,c}^1s{c_{i,{\pi _c}}}},\, {q_{i,c}^2p{c_i}} \right],
\end{align}
where $\pi_c \in \mathcal{Q}$ is the sensor used in contributing $raw_c$, and $sc_{i,\pi_c}$ is the sensing capacity of CAV $i$ on type-$\pi_c$ sensor. Here, ${\mathbf{q}_{i,c}} = {\left[{q_{i,c}^1}, {q_{i,c}^2} \right]}$ represents the quality of content service (QoCS) of CAV $i$ in sharing content $c$. In specific, ${q_{i,c}^1}$ means the ratio of utilized sensing resource in contributing $raw_c$ to the total sensing capacity of type-$\pi_c$ sensor of CAV $i$, and ${q_{i,c}^2}$ indicates the ratio of utilized computing resource in generating $result_c$ to CAV $i$'s overall processing capacity $p{c_i}$.

Different content can have different popularity degrees. Here, the popularity distribution among all published contents in CAV fleet $k$ is denoted by ${\bf{f}}_k = \left[f_1,\cdots,f_c,\cdots,f_{C_k}\right]$, {which} can be modeled by the Zipf distribution \cite{8647913}, i.e.,
\begin{align}\label{eq:content}
{f_c} = \frac{1}{{{\left( {\tau \left( {c} \right)} \right)}^{ \kappa }}{\sum\nolimits_{l = 1}^{C_k} {{l^{ - \kappa }}} }},
\end{align}
where ${\tau \left( {c} \right)} \in [1, C_k]$ is the index of content $c$ with the decreasing order of the number of subscriber CAVs among all contents in the set $\mathcal{C}_k$. $\kappa$ is a positive value to characterize the content popularity. If $\kappa = 0$, it means that the popularity of contents follows the uniform distribution. The larger $\kappa$ implies that fewer popular contents account for the majority of requests.
Eq. (\ref{eq:content}) also indicates that a content with a larger ${\tau \left( {c} \right)}$ corresponds to a smaller popularity degree.

\subsection{Publish/Subscribe Model}\label{subsec:pubsubmodel}
The metadata of each published content $c \in \mathcal{C}_k$ in each topic $o \in \mathcal{O}_k$ is elaborated as:
\begin{align}\label{eq:1-4}
meta_c &= \left\langle {ID_i, t{_c}, \pi_c, MulticastAdd_c, } \right. \nonumber \\
&\left. {H({ra{w_c}}), H({{result_{c}}}), H(met{a_c}), Sig_i} \right\rangle,
\end{align}
where $ID_i$ is the unique identity (i.e., public key) of publisher CAV $i$ who publishes content $c$, $t{_c}$ is the publish time of content $c$, $\pi_c$ is the type of sensor for generating $raw_c$, $MulticastAdd_c$ is the multicast group address \cite{Akk2016multicast} assigned by master CAV $h_k$ to multicast content $c$ from publisher CAV $i$ to all subscriber CAVs in the set $\mathcal{J}_c$. $H\left( . \right)$ is the secure hash function, and $Sig_i$ is the signature of publisher CAV $i$ on message $H\left( met{a_c} \right)$.


\subsection{Security Model}\label{subsec:securitymodel}
In our security model, we define the following two kinds of attacks that may threaten CAVs' security during content transmission in the pub/sub framework.

\emph{1) Meaningless and Low-quality Content Publishing Attack.} Due to the selfishness and autonomy of CAVs' nature, {free-riding} CAVs may publish meaningless and low-quality vehicular content to save cost if there is not sufficient compensation for their costs in contributing contents, which may impede the spread of safety-critical information among CAVs.

\emph{2) Dishonest and Harmful Content Publishing Attack.} Malicious CAVs may publish false and even harmful vehicular content (e.g., injected with malwares and viruses) to multiple subscribers to threaten the normal driving activities of nearby CAVs and gain benefits. For instance, a malicious publisher CAV may disseminate forged traffic congestion messages to keep the road open for itself.

\section{Vehicular Pub/Sub Game}\label{sec:SPAD}
\begin{figure}[t]\setlength{\abovecaptionskip}{-0.cm}
\centering
  \includegraphics[width=7.0cm,height = 2.7cm]{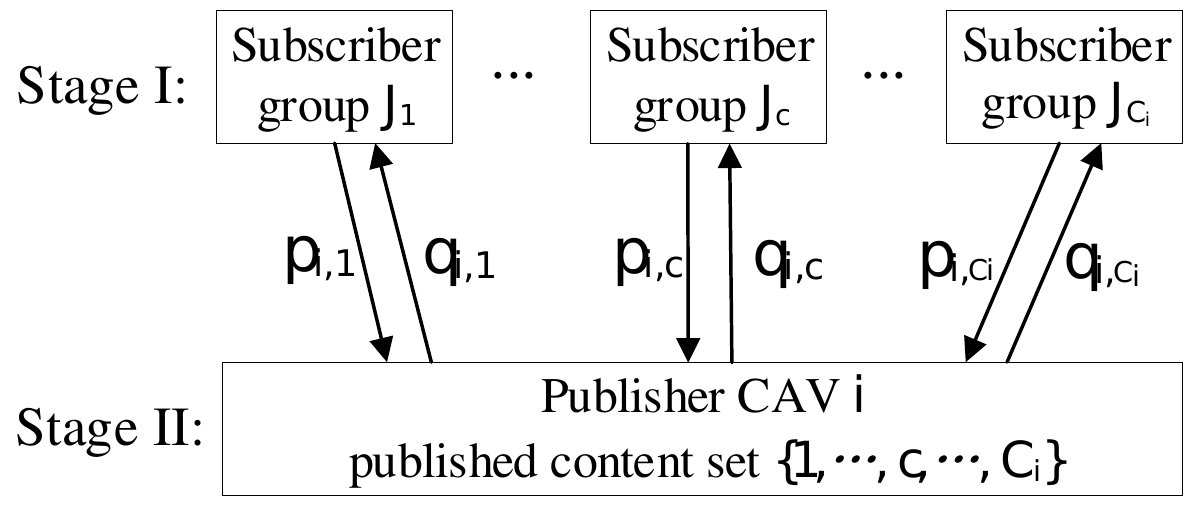}\\
  \caption{Structure of the proposed vehicular pub/sub game model.}\label{fig:stackelbergmodel}\vspace{-0.15cm}
\end{figure}

For motivating publisher CAVs to deliver high-quality content, each subscriber CAV $j \in \mathcal{J}_c$ with content preference $\beta_{j,c}$ selects the payment strategy ${\mathbf{p}_{i,j,c}} = {\left[{p_{i,j,c}^1},\, {p_{i,j,c}^2} \right]}$ to compensate the publisher CAV $i \in \mathcal{I}_k$ for the cost in contributing content $c \in \mathcal{C}_i$. Here, ${p_{i,j,c}^1}$ and ${p_{i,j,c}^2}$ are payments for acquiring the raw sensing data $raw_c$ and the processed result $result_c$ of content $c$, respectively.
Given payment strategies, each publisher CAV $i$ determines the QoCS strategy, i.e., ${\mathbf{q}_{i,c}} = {\left[{q_{i,c}^1},\,{q_{i,c}^2} \right]}$, on contributing content $c$. The criteria for evaluating the QoCS of content are given by the system and are known for all CAVs. {In the process of pub/sub}, each subscriber CAV desires high-quality content services with low payment, whereas each publisher CAV hopes the payment can be as high as possible. Accordingly, a competition exists between the publisher CAV and every subscriber CAV.
Based on \cite{Liu2019-2}, the competitive interactions between publisher CAV and subscriber CAVs can be formulated as a two-stage Stackelberg game{, as shown in Fig. \ref{fig:stackelbergmodel}}. 
In the game, all players are assumed to be rational and selfish, whose targets are to maximize their utilities.
To analyze the optimal strategy of each player, the utility functions of subscriber CAVs and publisher CAV need to be designed, respectively.

\subsection{Utility function of subscriber CAVs}\label{subsec:utilitysub}
Since there is only one copy of the published content $c =\{raw_c,result_c\}$ in topic $o$ to be transmitted to multiple subscriber CAVs in the set ${\mathcal{J}_c}$ through V2V multicasting \cite{Akk2016multicast}, the QoCS of content $c$ provided by publisher CAV $i$ for all subscriber CAVs with the same $\beta_{j,c}$ are identical.
Therefore, the payment vectors for each published content $c$ are assumed to be the same, i.e., $\mathbf{p}_{i,c} = {\left[ {p_{i,c}^1},\,{p_{i,c}^2} \right]} = {\mathbf{p}_{i,j,c}}, \forall j \in \mathcal{J}_c$. For ease of expression, we define ${\lambda _{j,c}} = \left[{\beta _{j,c}}, {1 - {\beta _{j,c}}} \right]^\mathbb{T}$, where $\mathbb{T}$ is the transpose symbol.
The utility function of subscriber CAV $j\in \mathcal{J}_c$ with content preference $\beta_{j,c}$ on consuming content $c$ can be defined as the difference between its satisfaction and cost (i.e., payment and transmission delay) in each content service. As such, we have 
\begin{align}\label{eq:utility-sub}
{\mathcal{U}_j}\left({\mathbf{q}_{i,c}},{\mathbf{p}_{i,c}}\right)& = \Omega \left({\mathbf{q}_{i,c}}\right) - {\left[ {\begin{array}{*{20}{c}}
{\vartheta _c^1\,{\beta _{j,c}}}\\
{\vartheta _c^2\left( {1 - {\beta _{j,c}}} \right)}
\end{array}} \right]^\mathbb{T}} \times \\ \nonumber
& {({\mathbf{p}_{i,c}})^\mathbb{T}}{\mathbf{q}_{i,c}}{\lambda _{j,c}} - {\left[ {\begin{array}{*{20}{c}}
{{\gamma ^1}{\beta _{j,c}}}\\
{{\gamma ^2}\left( {1 - {\beta _{j,c}}} \right)}
\end{array}} \right]^\mathbb{T}}{\bf{T}}_{i,j,c}^{delay},
\end{align}
where ${q_{i,c}^1},{q_{i,c}^2} \in [0, 1]$. For any $u \in \{1,2\}$, ${q_{i,c}^u} = 1$ implies the highest QoCS, while ${q_{i,c}^u} = 0$ means that the QoCS is the lowest. $\Omega({\mathbf{q}_{i,c}})$ is the satisfaction function of subscriber CAV $j$ with the QoCS vector ${\mathbf{q}_{i,c}}$. ${\vartheta_c^1},{\vartheta_c^2}$ are positive price adjustment parameters. ${\gamma^1},{\gamma^2}$ are positive adjustment factors. ${\bf{T}}_{i,j,c}^{delay}$ is the transmission delay vector of content $c$ and is calculated by
\begin{align}
{\bf{T}}_{i,j,c}^{delay} = \frac{1}{{{r_{i,j}}}}{\left[{s_c^1},{s_c^2}\right]^\mathbb{T}} ,
\end{align}
where ${s_c^1},{s_c^2}$ are the data size of $raw_c$ and $result_c$, respectively. ${r_{i,j}}$ is the transmission rate between publisher CAV $i$ and subscriber CAV $j\in\mathcal{J}_c$. {Full-duplex V2V multicasting communications \cite{9489369,7968385} are employed for each CAV to support simultaneous data publications and/or subscriptions over the same or different subchannels.
The path-loss of V2V link between CAV $i$ and CAV $j$ can be described by the large scale fading channel model \cite{7968385}, and the channel gain is given by
\begin{align}
\Phi _{i,j} = |\mu_0|^2(d_{i,j})^{-\Im},
\end{align}
where $\mu_0$ is the Rayliegh fading channel coefficient, $\Im>0$ is the path-loss exponent of a V2V link, and $d_{i,j}$ is the Euclidean distance between two CAVs $i$ and $j$. All channels are supposed to be reciprocal \cite{7005537}, i.e., $\Phi _{i,j}=\Phi _{j,i}$.
According to \cite{7968385}, to ensure a desirable SINR threshold $\mathrm{SINR}_\lambda$ at subscriber CAV $j$, the transmit power of publisher CAV $i$ over an intra-fleet V2V link satisfies:
\begin{align}
P_i^{Tr} = P_i^{0} + \frac{\mathrm{SINR}_\lambda }{|\mu_0|^2}{\cdot \Re\cdot (d_{i,j})^{\Im}},
\end{align}
where $P_i^{Tr}$ is the transmit power of CAV $j$ in the presence of both additive white Gaussian noise (AWGN) and co-subchannel interference. $P_i^{0}$ is CAV $j$'s transmit power with the presence of AWGN. $\Re$ is the interference of using the same subchannel\footnotemark[3]. Hence, we have $r_{i,j} = B_i^{Tr}{\log _2}\left( 1 + \mathrm{SINR}_\lambda \right)$, where $B_i^{Tr}$ is the subchannel bandwidth.}
\footnotetext[3]{{The self-interference of CAVs in data transmission over the same subchannel can be efficiently mitigated via existing self-interference cancellation (SIC) models \cite{9489369} and subchannel allocation models \cite{7968385}.}}

Based on \cite{Liu2019-2,9288629}, the natural logarithmic function, which is widely adopted in modeling the utilities of content consumers, is employed to model the satisfaction function $\Omega({\mathbf{q}_{i,c}})$. We have 
\begin{align}
\Omega ({{\bf{q}}_{i,c}}) &= {\alpha _{i,c}}{f_c}{R_i}\log \left( {1 + {\mathbf{Q}_{i,c}}{\lambda _{j,c}}} \right) \nonumber \\
&={\alpha _{i,c}}{f_c}{R_i}\log \left( {1 + {{\left[ {\begin{array}{*{20}{c}}
{sc_{i,\pi_c}\,q_{i,c}^1}\\
{p{c_i}\,q_{i,c}^2}
\end{array}} \right]}^\mathbb{T}}{\lambda _{j,c}}} \right),
\end{align}
where ${\alpha _{i,c}} > 0$ is the satisfaction coefficient of subscriber CAVs on content $c$, ${f_c}$ is the popularity degree of content $c$ defined in Eq. (\ref{eq:content}), and ${R_i}$ is the reputation value of publisher CAV $i$ defined in {Sect.~\ref{subsec:reputation}}.

Accumulating utilities ${\mathcal{U}_j}\left({\mathbf{q}_{i,c}},{\mathbf{p}_{i,c}}\right)$ over all $j\in \mathcal{J}_c$, the overall utility function of subscriber CAV group $\mathcal{J}_c$ can be obtained as:
\begin{align}\label{eq:utility-subg}
{\mathcal{U}_{{{\cal J}_c}}}&\left( {{{\bf{q}}_{i,c}},{{\bf{p}}_{i,c}}} \right) = \sum\nolimits_{j \in {{\cal J}_c}} {{u_j}\left( {{{\bf{q}}_{i,c}},{{\bf{p}}_{i,c}}} \right)}  \nonumber  \\
=\ &{\alpha _{i,c}}{f_c}{R_i}\sum\nolimits_{j \in {{\cal J}_c}} {\log \left( {1 + {{\left[ {\begin{array}{*{20}{c}}{s{c_{i,\pi_c}}\,q_{i,c}^1}\\ {p{c_i}\,q_{i,c}^2}\end{array}} \right]}^\mathbb{T}}{\lambda _{j,c}}} \right)} \nonumber \\
& - \sum\nolimits_{j \in {{\cal J}_c}} {{{\left[ {\begin{array}{*{20}{c}}
{\vartheta _c^1\, p_{i,c}^1 q_{i,c}^1}\\
{\vartheta _c^2\, p_{i,c}^2 q_{i,c}^2}
\end{array}} \right]}^\mathbb{T}}{\lambda _{j,c}}}  - \sum\nolimits_{j \in {{\cal J}_c}} {{{\left[ {\begin{array}{*{20}{c}}
{{\gamma ^1}s_c^1}\\
{{\gamma ^2}s_c^2}
\end{array}} \right]}^\mathbb{T}}}  \nonumber \\
&\times{\lambda _{j,c}} \left(B_i^{Tr}{\log _2}\left( 1 + \mathrm{SINR}_\lambda \right)\right)^{-1}.
\end{align}

\subsection{Utility function of publisher CAVs}\label{subsec:utilitypub}
The utility function of publisher CAV $i \in \mathcal{I}_k$ on contributing contents $\{1, \cdots, C_i\}$ can be defined as the revenue minus its cost in all content services, which can be expressed as:
\begin{align}\label{eq:utility-pub}
{\mathcal{U}_i}\left( {{{\bf{q}}_i},{{\bf{p}}_i}} \right) &= \sum\nolimits_{c \in {{\cal C}_i}}\left( { \sum\nolimits_{j \in {{\cal J}_c}} {{{\left[ {\begin{array}{*{20}{c}}
{\vartheta _c^1\, p_{i,c}^1 q_{i,c}^1}\\  {\vartheta _c^2\, p_{i,c}^2 q_{i,c}^2} \end{array}} \right]}^\mathbb{T}}{\lambda _{j,c}}} } \right. \\ \nonumber
& \left. { - {\phi _{i,c}}({{\bf{q}}_{i,c}}) - \mathbf{I} \times \phi _c^{energy} - {\phi_0}  } \right),
\end{align}
where ${{\bf{q}}_{i}} = \left[{\mathbf{q}_{i,1}}, \cdots, {\mathbf{q}_{i,c}}, \cdots, {\mathbf{q}_{i,{C}_i}}\right]$ is the QoCS vector of publisher CAV $i$, ${\mathbf{p}_i} = \left[{\mathbf{p}_{i,1}}, \cdots, {\mathbf{p}_{i,c}}, \cdots, {\mathbf{p}_{i,{C}_i}}\right]$ is the payment vector of all subscriber CAV groups on content set $\mathcal{C}_i$, and $\phi_{i,c}({\mathbf{q}_{i,c}})$ is the content generation cost of publisher CAV $i$ with QoCS ${\mathbf{q}_{i,c}}$. $\mathbf{I} = [1_{\{\mathcal{J}_c^1 \ne \emptyset\}},\,1_{\{\mathcal{J}_c^2 \ne \emptyset\}}]$ is a vector, where $1_x$ is an indicator function if the event $x$ it true; otherwise it equals to zero. $\phi _c^{energy}$ is the energy consumption cost in transmitting content $c$ and is calculated by
\begin{align}
\phi _c^{energy} = {P_i^{Tr}}\,{\bf{T}}_{i,j,c}^{delay} .
\end{align}

{In Eq. (\ref{eq:utility-pub}),} ${\phi_0}$ is the content {management} fee paid to master CAV $h_k$ for each published content. Here, the cost in contributing content $c$, i.e., $\phi_{i,c}({\mathbf{q}_{i,c}})$, is associated with the type of sensor, the sensing or processing capacity, and the QoCS of publisher CAV.
Based on the quadratic cost model \cite{Wu2011a}, $\phi_{i,c}({\mathbf{q}_{i,c}})$ can be formulated as the quadratic function of the service quality, i.e.,
\begin{align}
\phi_{i,c}({\mathbf{q}_{i,c}}) = \mathbf{I} \times {\left[ {\begin{array}{*{20}{c}}
{\xi _c^1\, \varepsilon _{i,\pi_c}^1  sc_{i,\pi_c} \cdot{{\left( {q_{i,c}^1} \right)}^2}}\\
{\xi _c^2\,  \varepsilon _i^2\, pc_i \cdot{{\left( {q_{i,c}^2} \right)}^2}}
\end{array}} \right]},
\end{align}
where $\xi _c^1,\xi _c^2$ are positive adjustment coefficients, $\varepsilon _{i,\pi_c}^1$ is the cost parameter of publisher CAV $i$ with the highest QoCS in contributing $raw_c$ using type-$\pi_c$ sensor, and $\varepsilon _i^2$ is the cost parameter of publisher CAV $i$ with the highest QoCS in computing $result_c$.

\subsection{Reputation Evaluation}\label{subsec:reputation}
The reputation value ${R_n}$ of each authorized CAV node $n \in \mathcal{N}$ is constructed from the {combination of vehicle's social role dimension and vehicle's} behavior dimension. We have
\begin{align}\label{eq:reputation}
{R_n} = {\lambda _R} f(role_n) + {\lambda _B} f\left( {behavio{r_n}} \right),
\end{align}
where ${\lambda _R}$ and ${\lambda _B}$ are positive normalization coefficients to guarantee that $R_n \in [0,1]$. $f(role_n)$ and $f\left( {behavio{r_n}} \right)$ are {role} and behavior effects of CAV $n$, respectively. Commonly, CAVs with different social role types {hold} different levels of trustworthiness, e.g., the police cars are more trustworthy than private cars.
Let $\mathcal{A} = \{1,\cdots,a,\cdots,A\}$ denote the set of registered vehicle role categories of CAVs, e.g., police cars, ambulances, private cars, etc. When a new CAV $n$ joins the network, it needs to submit its type of social role (i.e., $a \in \mathcal{A}$) by binding with its real identity (e.g., vehicle license number) in the registration phase at the {trusted authority (TA), such as the} certificate authority.
Here, the effect of CAV $n$'s vehicle role in reputation assessment is formulated as:
\begin{align}\label{eq:}
f\left( {rol{e_n}} \right) = {\mathbf{b}_{n}}  \cdot \mathbf{{v}}^\mathbb{T} = \sum\nolimits_{a \in \mathcal{A}} {{b_{n,a}}}  \cdot {v}_{a} ,
\end{align}
where ${\mathbf{b}_{n}} \!=\! [b_{n,1}, \cdots, b_{n,a}, \cdots, b_{n,A}]$ is the binary social role vector of CAV $n$. If $b_{n,a} = 1$, it means that the registered social role of CAV $n$ at TA is $a \in \mathcal{A}$. Otherwise, $b_{n,a} = 0$. $\mathbf{{v}} = [{v}_{1}, \cdots, {v}_{a}, \cdots, {v}_{A}]$ is the trustworthiness degree vector of CAVs with different types of registered vehicle roles.

Subscriber CAVs who doubt about the authenticity of the processed results $result_c$ or receive harmful sensory data $raw_c$ can report the corresponding publisher CAV $i \in \mathcal{I}_k$ to the master CAV $h_k$.
Then, master CAV $h_k$ collects the evidence by acquiring the raw sensory data $raw_c$ of content $c$ from the publisher CAV $i$, and decides whether publisher CAV $i$ misbehaves or not via digital forensics. The detailed forensics procedure for source identification and evidence collection can refer to \cite{Peng2018image,Chen2015video}.
The behavior effect $f({behavio{r_n}})$ of CAV $n$ in reputation calculation is related to the positive behavior effect (i.e., ${\digamma_n^P}$) and negative behavior effect (i.e., ${\digamma_n^N}$).
The positive behavior effect of CAV $n$ can be expressed as:
\begin{align}\label{eq:positiveBehave}
{\digamma_n^P} = {w_1}\sum\nolimits_{b = 1}^{N_n^{report}} {{e^{ - {\eta _1}\left( {t - {t_b}} \right)}}}  + {w_2}T_n^{recent},
\end{align}
where ${w_1}$ and ${w_2}$ are positive adjustment parameters, ${N_n^{report}}$ is the number of successful report times of CAV $n$, and ${T_n^{recent}}$ is the duration without misbehavior in recent time of CAV $n$, i.e., time interval from the occurrence time of the latest recorded misbehavior till the current time $t$. ${{e^{(.)}}}$ is the time decay function to describe the feature that latest behaviors are more important than older ones, ${\eta _1}>0$ is the decay factor, and $t_b$ is the occurrence time of $b$-th behavior. By considering time fading effects, the impact of old behavior records can be gradually reduced.
The negative behavior effect of CAV $n$ is associated with its number of recorded malicious behaviors and the occurrence time of each misbehavior, i.e.,
\begin{align}\label{eq:negativeBehave}
{\digamma_n^N} = {w_3}\sum\nolimits_{b = 1}^{N_n^{mis}} {{e^{ - {\eta _2}\left( {t - {t_b}} \right)}}},
\end{align}
where ${w_3}$ is a positive adjustment parameter, ${N_n^{mis}}$ is the number of CAV $n$'s recorded misbehaviors, and ${\eta _2}>0$ is the decay factor.

Based on the {standard} Bayesian inference model \cite{Maga2019}, the behavior effect can be defined in the form of beta distribution, i.e., $f({behavio{r_n}}) \sim beta(\alpha_n, \beta_n)$, where $\alpha_n = {\digamma_n^P} + 1$ and $\beta_n = {\digamma_n^N} + 1$. The beta probability density function (PDF) $g( \theta |\alpha, \beta)$ can be expressed by using a gamma function $\Gamma$ as:
\begin{align}\label{eq:beta}
g(\theta|\alpha ,\beta ) \!=\! \frac{{{\theta^{\alpha - 1}}{{\left( {1 - \theta} \right)}^{\beta - 1}}}}{{\int_0^1 {{\mu ^{\alpha \!-\! 1}}{{\left( {1 \!-\! \mu} \right)}^{\beta \!-\! 1}}d\mu } }} \!=\! \frac{{\Gamma(\alpha  + \beta )}}{{\Gamma(\alpha )\Gamma(\beta )}}{\theta^{\alpha  - 1}}{\left( {1 \!-\! \theta} \right)^{\beta - 1}},
\end{align}
where $0 \leq \theta \leq 1$.
Then, $f({behavio{r_n}})$ can be formulated as the expectation of beta PDF, i.e.,
\begin{align}\label{eq:bayes-e}
f({behavio{r_n}}) = \mathbb{E}\left[ {g(\theta|\alpha_n ,\beta_n )} \right] = \frac{\alpha _n}{{\alpha _n} + {\beta _n}}.
\end{align}
Initially, due to the absence of direct observations, each CAV has the same prior knowledge $beta(1, 1)$. {For improved accuracy and robustness in reputation evaluation, we devise an improved Bayesian inference model.}
Commonly, the reputation of each CAV node grows slowly, whereas it can be destroyed quickly when CAV misbehaves. Here, a punishment factor $\gamma > 1$ is introduced to punish CAVs' misbehaviors with the following advantages.
On one hand, it can significantly lower the reputation value of a malicious CAV if its misbehavior is detected. On the other hand, the reputation value of a malicious CAV will recover slowly even if it behaves legitimately in next time slots.
The modified behavior effect of CAV $n$ in Eq. (\ref{eq:bayes-e}) can be rewritten as:
\begin{align}\label{eq:}
f({behavio{r_n}}) = \frac{\alpha _n}{{\alpha _n} + \gamma {\beta _n}}.
\end{align}

{Besides, in contrast to the standard Bayesian model which assigns the same weight regardless of occurrence time of the observed behaviors, a CAV may change its behavior over time owing to the network dynamics. A time fading mechanism is adopted to address this issue by allowing CAVs to gradually forget old observations and assigning higher weights for recent observations, as shown in Eqs. (\ref{eq:positiveBehave})--(\ref{eq:negativeBehave}). Here, ${\eta _1},{\eta _2}$ are decay factors for CAVs' positive behavior effect and negative behavior effect, respectively.} 


\subsection{Optimization Problems}\label{subsec:problem}
The proposed vehicular pub/sub game between the publisher CAV and subscriber CAVs in content transmission can be formulated as:
\begin{align}\label{eq:}
\mathbb{G} = \left\{ {\left( {i,{\mathcal{J}_{1 \le c \le {{C}_i} }}} \right);\left( {{{\bf{q}}_i},{{\bf{p}}_i}} \right);\left( {{\mathcal{U}_i},{\mathcal{U}_{{{\cal J}_{1 \le c \le  {{C}_i} }}}}} \right)} \right\},
\end{align}
which can be regarded as a multiple-leaders and one-follower Stackelberg game.
As shown in Fig. \ref{fig:stackelbergmodel}, each subscriber CAV group $\mathcal{J}_c$, as the leader of the game $\mathbb{G}$, first chooses its optimal payment strategy ${\mathbf{p}_{i,c}}^* $ in stage I to maximize its utility defined in Eq. (\ref{eq:utility-subg}). Then the publisher CAV $i$, as the follower of the game, determines its optimal QoCS strategy ${{{\bf{q}}_{i}}}^*$ in stage II to maximize its utility defined in Eq. (\ref{eq:utility-pub}). As such, two optimization problems are formulated.

\emph{\textbf{Problem 1.}} The objective of each subscriber CAV group $\mathcal{J}_c$ is to maximize its utility function by determining its payment strategy ${\mathbf{p}_{i,c}}$, whereby the optimization problem $\mathcal{P}_1$ is:
\begin{align}\label{eq:p1}
\begin{array}{l}
\mathcal{P}_1:~ \mathop {\max }\limits_{{\mathbf{p}_{i,c}}}\, {\mathcal{U}_{{{\cal J}_c}}}\left( {{{\bf{q}}_{i,c}},{{\bf{p}}_{i,c}}} \right), ~ \forall \mathcal{J}_c \subseteq \mathcal{J}_k, \\[0.3cm]
{s.t.\left\{ \begin{array}{l}
0\le p_{i,c}^u \le p_{\max}, ~~\forall u \in \left\{ {1,2} \right\}, \forall c \in \mathcal{C}_i,\\[0.1cm]
{R_i} \ge \theta_{{\mathcal{J}_c}},~~~~~~~~~~\forall i \in {\mathcal{I}_k},
\end{array} \right.}
\end{array}
\end{align}
where $p_{\max}$ is the price budget of subscriber CAVs which indicates the highest payment in a content service, and $\theta_{{\mathcal{J}_c}}$ is the reputation threshold of subscriber CAV group ${\mathcal{J}_c}$.

\emph{\textbf{Problem 2.}} The objective of each publisher CAV $i \in \mathcal{I}_k$ is to maximize its utility function by selecting its QoCS strategy ${\mathbf{q}_{i}}$, whereby the optimization problem $\mathcal{P}_2$ is:
\begin{align}\label{eq:p2}
\begin{array}{l}
\mathcal{P}_2: \mathop {\max }\limits_{{\mathbf{q}_{i,1}},{\mathbf{q}_{i,2}},\cdots,{\mathbf{q}_{i,C_i}}} {\mathcal{U}_i}\left( {{{\bf{q}}_i},{{\bf{p}}_i}} \right),~ \forall i \in \mathcal{I}_k, \\[0.3cm]
s.t.~\, 0\le q_{i,c}^u \le 1, ~~~ \forall u \in \left\{ {1,2} \right\}, \forall c \in \mathcal{C}_i.
\end{array}
\end{align}

The solution of the game $\mathbb{G}$ is to find the Stackelberg equilibrium (SE), from which neither the publisher CAV nor the subscriber CAVs can deviate to improve their utilities.

\emph{\textbf{Definition 1.}}
The SE of game $\mathbb{G}$ is denoted by $\left({{\bf{p}}_i}^*, {{\bf{q}}_i}^*\right)$, where ${{\bf{p}}_i}^* = \left[ {\mathbf{p}_{i,c}}^*\right]_{1\leq c \leq {C_i}}$ is the solution for $\mathcal{P}_1$, and ${{\bf{q}}_i}^* = \left[ {\mathbf{q}_{i,c}}^*\right]_{1\leq c \leq {C_i}}$ is the solution for $\mathcal{P}_2$. Here, ${{\bf{p}}_{i,c}}^* = \left[ {p{{_{i,c}^1}^*},p{{_{i,c}^2}^*}} \right]$ and ${{\bf{q}}_{i,c}}^* = \left[ {q{{_{i,c}^1}^*},q{{_{i,c}^2}^*}} \right]$. Then a SE of the proposed game (if one exists) can be given by:\vspace{0.1mm}
\begin{numcases}{}
{\mathbf{p}_{i,c}}^* = \arg {{\max }_{{\mathbf{p}_{i,c}}}}\,{\mathcal{U}_{{{\cal J}_c}}}\left( {{\mathbf{q}_{i,c}}^*,{\mathbf{p}_{i,c}}} \right),  \forall {{\cal J}_c} \subseteq {{\cal J}_k},\\[0.03cm]
\ {{\bf{q}}_i}^* ~= \arg {{\max }_{{{\bf{q}}_i}}}\,{\mathcal{U}_i}\left( {{{\bf{q}}_i},{{\bf{p}}_i}^*} \right), ~~~~~~\; \forall i \in {{\cal I}_k}.
\end{numcases}\vspace{0mm}

\section{Static Vehicular Pub/Sub Game Analysis}\label{sec:game}
In this section, we analyze the SE of the static vehicular pub/sub game with one interaction to attain the optimal strategies of both publisher CAVs and subscriber CAVs. Here, all the parameters in the game model (e.g., satisfaction coefficient, cost parameter, and sensing/processing capacity) are public knowledge to all players.
To obtain the SE, the backward induction approach \cite{Liu2019-2} is exploited, where the optimal strategy of the follower (i.e., publisher CAV) is first analyzed followed by the optimal strategy analysis of the leader (i.e., subscriber CAV group).

\subsection{Optimal Strategy of Publisher CAV}\label{subsec:optimalq}
In stage II, publisher CAV $i$ decides its optimal QoCS strategy ${{\bf{q}}_{i}}^*$ to maximize its utility based on Theorem 1. 

\emph{\textbf{Theorem 1.}}
The optimal QoCS strategies of publisher CAV $i$ on contributing the sensing data $raw_c$ and the processing result $result_c$ of content $c \in \mathcal{C}_i$ are
\begin{align}\label{eq:q*1}
{{q}_{i,c}^1}^* = \left\{ \begin{array}{ll}
1,&if\ \resizebox{.21\hsize}{!}{$\frac{{2{\xi_c^1} {\varepsilon _{i,\pi_c}^1}{sc_{i,\pi_c}}}}{{J_c^1{\vartheta_c^1} }}$} \le {p_{i, c}^1} \le {p_{\max}};\\[0.1cm]
\resizebox{.23\hsize}{!}{$\frac{{J_c^1\,{\vartheta_c^1}\,{p_{i,c}^1}}}{{2{\xi_c^1} {\varepsilon _{i,\pi_c}^1}{sc_{i,\pi_c}}}}$},&if\ 0 < {p_{i,c}^1} < \resizebox{.21\hsize}{!}{$\frac{{2{\xi_c^1} {\varepsilon _{i,\pi_c}^1}{sc_{i,\pi_c}}}}{{J_c^1{\vartheta_c^1} }}$}.
\end{array} \right.
\end{align}
\begin{align}\label{eq:q*2}
{{q}_{i,c}^2}^* = \left\{ \begin{array}{ll}
1,&if\ \resizebox{.145\hsize}{!}{$\frac{{2{\xi_c^2} {\varepsilon _i^2}\,{pc_i}}}{{J_c^2{\vartheta_c^2} }}$} \le {p_{i, c}^2} \le {p_{\max}};\\[0.1cm]
\resizebox{.16\hsize}{!}{$\frac{{J_c^2\,{\vartheta_c^2}\,{p_{i,c}^2}}}{{2{\xi_c^2} {\varepsilon _i^2}\,{pc_i}}}$},&if\ 0 < {p_{i,c}^2} < \frac{{2{\xi_c^2} {\varepsilon _i^2}\,{pc_i}}}{{J_c^2{\vartheta_c^2} }}.
\end{array} \right.
\end{align}

\begin{IEEEproof}
{Please refer to Appendix \ref{Appendix A}.}
\end{IEEEproof}

\subsection{Optimal Strategy of Subscriber CAV}\label{subsec:optimalp}
In stage I, given the optimal QoCS strategy ${\mathbf{q}_{i,c}}^*$ in Eqs. (\ref{eq:q*1})--(\ref{eq:q*2}), each subscriber CAV group $\mathcal{J}_c$ decides its optimal payment strategy ${{\mathbf{p}_{i,c}}^*}$ for content $c$ to maximize its utility ${\mathcal{U}_{{\mathcal{J}_c}}}\left( {{\mathbf{q}_{i,c}}^*,{\mathbf{p}_{i,c}}} \right) $ according to the following theorem.

\emph{\textbf{Theorem 2.}}
The optimal payment strategies of subscriber CAV group $\mathcal{J}_c$ for the sensing data $raw_c$ and the processing result $result_c$ of content $c \in \mathcal{C}_i$ are
\begin{align}\label{eq:p*1}
{p_{i,c}^1}^* = \left\{ \begin{array}{ll}
\frac{2}{{J_c^1{\vartheta_c^1} }}{{\xi_c^1} {\varepsilon _{i,\pi_c}^1}{sc_{i,\pi_c}}}, &if\ \Psi^1 \ge 0;\\[0.15cm]
\frac{1}{{{J_c^1}{\vartheta_c^1} }}\left({\sqrt {\Upsilon^1}   - {\xi_c^1} {\varepsilon _{i,\pi_c}^1}}\right),&if\ \Psi^1 < 0.
\end{array} \right.
\end{align}
\begin{align}\label{eq:p*2}
{p_{i,c}^2}^* = \left\{ \begin{array}{ll}
\frac{2}{{J_c^2{\vartheta_c^2} }}{{\xi_c^2} {\varepsilon _i^2}\,{pc_i}}, &if\ \Psi^2 \ge 0;\\[0.15cm]
\frac{1}{{{J_c^2}{\vartheta_c^2} }}\left({\sqrt {\Upsilon^2}   - {\xi_c^2} {\varepsilon _i^2}}\right),&if\ \Psi^2 < 0 ,
\end{array} \right.
\end{align}
where
\begin{numcases}{}	
{\Upsilon ^1} = {\left( {\xi _c^1\varepsilon _{i,{\pi _c}}^1} \right)^2} + J_c^1{\alpha _{i,c}}{f_c}{R_i}\xi _c^1\varepsilon _{i,{\pi _c}}^1s{c_{i,{\pi _c}}}, \hfill \nonumber \\
{\Upsilon ^2} = {\left( {\xi _c^2\varepsilon _i^2} \right)^2} + J_c^2{\alpha _{i,c}}{f_c}{R_i}\xi _c^2\varepsilon _i^2p{c_i}, \hfill \nonumber \\
{\Psi ^1} = J_c^1{\alpha _{i,c}}{f_c}{R_i} - 4\xi _c^1\varepsilon _{i,{\pi _c}}^1\left( {s{c_{i,{\pi _c}}} + 1} \right), \hfill \nonumber \\
{\Psi ^2} = J_c^2{\alpha _{i,c}}{f_c}{R_i} - 4\xi _c^2\varepsilon _i^2\left( {p{c_i} + 1} \right). \nonumber \hfill
\end{numcases}

\begin{IEEEproof}
{Please refer to Appendix \ref{Appendix B}.}
\end{IEEEproof}

According to the above analysis, the SE of the proposed game can be attained, which is shown in the following theorem.

\emph{\textbf{Theorem 3.}}
The SE of the static vehicular pub/sub game $\mathbb{G}$ is given by:
\begin{align}\label{eq:stackSE}
\left( {{p_{i,c}^u}^*,{q_{i,c}^u}^*} \right) \!=\! \left\{ \begin{array}{ll}
\left(  \resizebox{.10\hsize}{!}{${\frac{{2}}{{{J_c^u}{\vartheta_c^u} }}}$}  {\Lambda^u},1 \right)\!,\!~~if\  \Psi^u  \!\ge\! 0 ;\\[0.2cm]
\left(\resizebox{.33\hsize}{!}{$ {\frac{{\sqrt {\Upsilon^u}   - \Omega^u}}{{{J_c^u}{\vartheta_c^u} }},\frac{{\sqrt {\Upsilon^u} - \Omega^u}}{2\Lambda^u}}$} \right)\!,\!\,\ if\ \Psi^u \!<\! 0 ,
\end{array} \right.
\end{align}
where $u \in \{1,2\}$, $\Omega^1 = {\xi_c^1}{\varepsilon_{i,{\pi _c}}^1}$, $\Omega^2 = {\xi_c^2}{\varepsilon_{i}^2}$, $\Lambda^1 = {\Omega^1}{sc_{i,{\pi _c}}}$, and $\Lambda^2 = {\Omega^2}{pc_{i}}$.

\begin{IEEEproof}
{Please refer to Appendix \ref{Appendix C}.}
\end{IEEEproof}

\emph{\textbf{Remark.}}
In the competitive pub/sub process, subscriber CAVs with different subscription preferences $\beta_{j,c}$'s and publisher CAVs with different sensing/processing capacities can determine their strategies {for each content service} by obeying the SE in Eq. (\ref{eq:stackSE}) to obtain their maximized utilities {in a distributed manner}. 

\section{Dynamic Vehicular Pub/Sub Game With Two-Tier Hotbooting PHC-based Learning}\label{sec:learning}
In this section, we analyze the dynamic vehicular pub/sub game with repeated interactions between publisher CAVs and subscriber CAVs.
Different from the static vehicular pub/sub game where the parameters of both publisher and subscriber CAVs' utility functions are public knowledge, in a practical network, these parameters are typically private and cannot be readily available for all participants. Alternatively, both publisher CAVs and subscriber CAVs can apply reinforcement learning techniques to search their optimal strategies via trial and error under multiple interactions. The strategy-making processes of both publisher CAVs and subscriber CAVs can be modeled as finite Markov decision processes (MDPs) in the dynamic game \cite{9159929,10106022}.

\subsection{Hotbooting PHC-Based Payment Strategy}\label{subsec:payment}
A high payment for requested data decreases the immediate utility of the subscriber CAV, whereas it stimulates the higher quality of contents shared by publisher CAV in the future. Hence, the current payment strategies of subscriber CAV groups influence the long-term quality of subscribed contents and their future payoffs.
By formulating the pricing decision-making process as a finite MDP, each subscriber CAV group can apply the hotbooting PHC (as an extension of Q-learning) to search its optimal payment strategy without explicitly knowing the private parameters of the publisher CAV's utility model.
At each time slot $t$, the system state vector ${\bf{z}}_{i,c}^t = \left[({{z}_{i,c}^1})^t,({{z}_{i,c}^2})^t \right]$ for subscriber CAV group ${{\cal J}_c}$ consists of the previous QoCS of publisher CAV $i$ in contributing content $c$, i.e., ${\bf{z}}_{i,c}^t = {\bf{q}}_{i,c}^{t-1}$. Here, $({{z}_{i,c}^u})^t = ({{q}_{i,c}^u})^{t-1}$, $\forall u \in \{1,2\}$. For simplicity, the payment strategy of each subscriber CAV group is uniformly quantized into $X+1$ levels, i.e., ${{p}_{i,c}^u} \in \mathcal{X}=\{\frac{x}{X}\cdot p_{\max}\}_{0 \le x \le X}$, $\forall u \in \{1,2\}$. In the learning process, the reward of subscriber CAV group ${{\cal J}_c}$ is defined as its scaled utility $\lambda_1 {\mathcal{U}_{{\cal J}_c}}\left({\bf{q}}_{i,c}^t,{\bf{p}}_{i,c}^t\right)$, where $\lambda_1$ is a positive scale factor.

Let $\mathbb{{Q}}\left({\bf{z}}_{i,c}^t, {\bf{p}}_{i,c}^t\right)$ denote the Q-function of state ${\bf{z}}_{i,c}^t$ and action ${\bf{p}}_{i,c}^t = \left[({{p}_{i,c}^1})^t,({{p}_{i,c}^2})^t \right]$, which indicates the expected long-term discounted utility of the state-action pair $\left({\bf{z}}_{i,c}^t, {\bf{p}}_{i,c}^t\right)$. The Q-function of subscriber CAV group ${{\cal J}_c}$ can be updated based on iterative Bellman equation as follows:
\begin{align}\label{eq:6-1}
\mathbb{{Q}}\left({\bf{z}}_{i,c}^t, {\bf{p}}_{i,c}^t\right) \leftarrow \mathbb{{Q}}&\left({\bf{z}}_{i,c}^t, {\bf{p}}_{i,c}^t\right) + \psi_1 \left\{ \lambda_1{\mathcal{U}_{{\cal J}_c}}\left({\bf{q}}_{i,c}^t,{\bf{p}}_{i,c}^t\right) \right. \nonumber \\
&\left. {+ \chi_1 \mathbb{{V}} \left({\bf{z}}_{i,c}^{t+1} \right) - \mathbb{{Q}}\left({\bf{z}}_{i,c}^t, {\bf{p}}_{i,c}^t\right)} \right\},
\end{align}
where $\psi_1 \in (0,1]$ is the learning rate implying the weight of current experience, and $\chi_1 \in [0,1]$ is the discount factor indicating the myopic view regarding the future reward. ${\bf{z}}_{i,c}^{t+1}$ is the new state vector of publisher CAV $i$ on content $c$ at time slot $t+1$, which is transformed from state ${\bf{z}}_{i,c}^{t}$ with action ${\bf{p}}_{i,c}^t$.
Besides, $\mathbb{{V}} \left({\bf{z}}_{i,c}^t \right)$ denotes the value function, which maximizes the Q-function at state ${\bf{z}}_{i,c}^t$ over the action set, i.e.,
\begin{align}\label{eq:6-2}
\mathbb{{V}} \left({\bf{z}}_{i,c}^{t+1} \right) \leftarrow \mathop {\max }\limits_{{\bf{p}}_{i,c}} \mathbb{{Q}}\left({\bf{z}}_{i,c}^{t+1}, {\bf{p}}_{i,c}^{t+1}\right).
\end{align}

To speed up the convergence time in traditional Q-learning, the proposed hotbooting PHC algorithm makes two improvements: 1) hotbooting preparation for efficient system initialization, and 2) PHC-based mixed-strategy table update for balancing the exploration and exploitation. In specific, in PHC, the mixed-strategy table $\pi\left({\bf{z}}_{i,c}^t, {\bf{p}}_{i,c}^t\right)$ is updated by increasing the probability of behaving greedily (i.e., opt the payment strategy with the highest Q-function) by a small value $\delta_1$, $0 < \delta_1 \leq 1$, and decreasing the other probabilities by $-\frac{\delta_1}{X}$. We have
\begin{align}\label{eq:6-3}
\pi &\left( \mathbf{z}_{i,c}^{t},\mathbf{p}_{i,c}^{t} \right) \gets \pi \left( \mathbf{z}_{i,c}^{t},\mathbf{p}_{i,c}^{t} \right)\nonumber \\
&+\left\{ \begin{array}{cl}
	\delta _1,&if\ \mathbf{p}_{i,c}^{*}=\arg\max_{\mathbf{p}_{i,c}}\mathbb{Q}\left( \mathbf{z}_{i,c}^{t},\mathbf{p}_{i,c} \right);\\[0.02cm]
	-\frac{\delta _1}{X+1},&otherwise.
\end{array} \right.
\end{align}
Then, subscriber CAV group ${{\cal J}_c}$ selects its payment strategy ${\bf{p}}_{i,c}^t$ based on the mixed-strategy table $\pi\left({\bf{z}}_{i,c}^t, {\bf{p}}_{i,c}^t\right)$, i.e.,
\begin{align}\label{eq:6-4}
\Pr\left( \mathbf{p}_{i,c}^{t} = \hat{\bf{p}}_{i,c} \right) = \pi \left( {\bf{z}}_{i,c}^t,\hat{\bf{p}}_{i,c} \right) ,\forall \hat{\mathbf{p}}_{i,c}\in \mathcal{X}.
\end{align}

To avoid inefficient random explorations in traditional Q-learning with all-zero Q-value initialization, a hotbooting technique (as shown in lines 7--13 in Algorithm 1) is utilized for subscriber CAV groups by exploiting the experience from similar scenarios to initialize the Q-value and mixed-strategy table. In specific, $W$ vehicular pub/sub experiments are conducted in similar scenarios before the game and each of them lasts $T$ time slots.
The output of Algorithm 1 through $W$ experiments (i.e., $\mathbb{{Q}}_h$ and $\pi_h$) is utilized as the input of Algorithm 2, with $\mathbb{{Q}} = \mathbb{{Q}}_h$ and $\pi = \pi_h$. The hotbooting PHC-based optimal payment strategy decision {process} for subscriber CAV groups is summarized in lines 6--13 in Algorithm \ref{Algorithm2}.

\begin{algorithm}[t!]\begin{small}
   \caption{\textbf{Hotbooting Preparation}}\label{Algorithm1}
    \begin{algorithmic}[1]
        \STATE \textbf{Input: }$\psi_1$, $\psi_2$, $\chi_1$, $\chi_2$, $\delta _1$, $\delta _2$, ${\bf{z}}_{i,c}^0$, $\tilde{\bf{z}}_{i,c}^0$, $X$, $Y$
        \STATE \textbf{Output: }${\mathbb{{Q}}}_h$, ${\pi}_h$, $\tilde{\mathbb{{Q}}}_h$, $\tilde{\pi}_h$
        \STATE \textbf{Initialize: }${\mathbb{{Q}}}_h=0$, ${\pi}_h=\frac{1}{X+1}$, $\tilde{\mathbb{{Q}}}_h=0$, $\tilde{\pi}_h=\frac{1}{Y+1}$
        \FOR {$w = 1, 2,\cdots,W $}
            \FOR {$t = 1, 2,\cdots,T $}
                \FOR {$c = 1, 2,\cdots,C_i $}
                    \STATE \COMMENT{run on the subscriber CAV group ${{\cal J}_c}$}
                    \STATE Set ${\bf{z}}_{i,c}^t = {\bf{q}}_{i,c}^{t-1}$.
                    \STATE Choose ${\bf{p}}_{i,c}^t$ via Eq. (\ref{eq:6-4}) and send it to publisher CAV $i$.
                    \STATE Observe and evaluate the QoCS ${\bf{q}}_{i,c}^t$.
                    \STATE Obtain utility ${\mathcal{U}_{{\cal J}_c}}\left({\bf{q}}_{i,c}^t,{\bf{p}}_{i,c}^t\right)$ via Eq. (\ref{eq:utility-subg}).
                    \STATE Update $\mathbb{{Q}}_h\left({\bf{z}}_{i,c}^t, {\bf{p}}_{i,c}^t\right)$ via Eqs. (\ref{eq:6-1}) and (\ref{eq:6-2}).
                    \STATE Update ${\pi}_h\left({\bf{z}}_{i,c}^t, {\bf{p}}_{i,c}^t\right)$ via Eq. (\ref{eq:6-3}).
                    \STATE \COMMENT{run on the publisher CAV $i$}
                    \STATE Set $\tilde{{\bf{z}}}_{i,c}^t = {\bf{p}}_{i,c}^{t-1}$.
                    \STATE Choose ${\bf{q}}_{i,c}^t$ via Eq. (\ref{eq:6-8}) and send it to subscriber CAV group ${{\cal J}_c}$.
                    \STATE Observe the payment ${\bf{p}}_{i,c}^t$.
                    \STATE Obtain utility ${\mathcal{U}_{i}}\left({\bf{q}}_{i,c}^t,{\bf{p}}_{i,c}^t\right)$ via Eq. (\ref{eq:utility-pub}).
                    \STATE Update $\tilde{\mathbb{{Q}}}_h\left(\tilde{{\bf{z}}}_{i,c}^t, {\bf{q}}_{i,c}^t\right)$ via Eqs. (\ref{eq:6-5}) and (\ref{eq:6-6}).
                    \STATE Update $\tilde{{\pi}}_h\left(\tilde{{\bf{z}}}_{i,c}^t, {\bf{q}}_{i,c}^t\right)$ via Eq. (\ref{eq:6-7}).
                \ENDFOR
            \ENDFOR
        \ENDFOR
    \end{algorithmic}\end{small}
\end{algorithm}

\begin{algorithm}[t!]\begin{small}
   \caption{\textbf{Two-Tier Hotbooting PHC-Based Optimal Strategy Decision}}\label{Algorithm2}
    \begin{algorithmic}[1]
    \STATE Run Algorithm \ref{Algorithm1} for hotbooting.
        \STATE \textbf{Initialize: }$\psi_1$, $\psi_2$, $\chi_1$, $\chi_2$, $\delta _1$, $\delta _2$, ${\bf{z}}_{i,c}^0$, $\tilde{\bf{z}}_{i,c}^0$, $X$, $Y$
        \STATE ${\mathbb{{Q}}} = {\mathbb{{Q}}}_h$, ${\mathbb{{V}}}=0$, ${\pi} = {\pi}_h$, $\tilde{{\mathbb{{Q}}}} = \tilde{{\mathbb{{Q}}}}_h$, $\tilde{{\mathbb{{V}}}}=0$, $\tilde{{\pi}} = \tilde{{\pi}}_h$
        \FOR {$t = 1, 2,\cdots,T $}
            \FOR {$c = 1, 2,\cdots,C_i $}
                \STATE \textbf{Tier 1: Hotbooting PHC-Based Payment Strategy.} \COMMENT{run on the subscriber CAV group ${{\cal J}_c}$}
                \STATE Set ${\bf{z}}_{i,c}^t = {\bf{q}}_{i,c}^{t-1}$.
                \STATE Choose ${\bf{p}}_{i,c}^t$ via Eq. (\ref{eq:6-4}) and send it to publisher CAV $i$.
                \STATE Observe and evaluate the QoCS ${\bf{q}}_{i,c}^t$.
                \STATE Obtain utility ${\mathcal{U}_{{\cal J}_c}}\left({\bf{q}}_{i,c}^t,{\bf{p}}_{i,c}^t\right)$ via Eq. (\ref{eq:utility-subg}).
                \STATE Update $\mathbb{{Q}}\left({\bf{z}}_{i,c}^t, {\bf{p}}_{i,c}^t\right)$ via Eq. (\ref{eq:6-1}).
                \STATE Update ${\mathbb{{V}}}\left({{\bf{z}}}_{i,c}^t\right)$ via Eq. (\ref{eq:6-2}).
                \STATE Update ${\pi}\left({\bf{z}}_{i,c}^t, {\bf{p}}_{i,c}^t\right)$ via Eq. (\ref{eq:6-3}).
                \STATE \textbf{Tier 2: Hotbooting PHC-Based QoCS Strategy.} \COMMENT{run on the publisher CAV $i$}
                \STATE Set $\tilde{{\bf{z}}}_{i,c}^t = {\bf{p}}_{i,c}^{t-1}$.
                \STATE Choose ${\bf{q}}_{i,c}^t$ via Eq. (\ref{eq:6-8}) and send it to subscriber CAV group ${{\cal J}_c}$.
                \STATE Observe the payment ${\bf{p}}_{i,c}^t$.
                \STATE Obtain utility ${\mathcal{U}_{i}}\left({\bf{q}}_{i,c}^t,{\bf{p}}_{i,c}^t\right)$ via Eq. (\ref{eq:utility-pub}).
                \STATE Update $\tilde{\mathbb{{Q}}}\left(\tilde{{\bf{z}}}_{i,c}^t, {\bf{q}}_{i,c}^t\right)$ via Eq. (\ref{eq:6-5}).
                \STATE Update $\tilde{\mathbb{{V}}}\left(\tilde{{\bf{z}}}_{i,c}^t\right)$ via Eq. (\ref{eq:6-6}).
                \STATE Update $\tilde{{\pi}}\left(\tilde{{\bf{z}}}_{i,c}^t, {\bf{q}}_{i,c}^t\right)$ via Eq. (\ref{eq:6-7}).
            \ENDFOR
        \ENDFOR
    \end{algorithmic}\end{small}
\end{algorithm}

\subsection{Hotbooting PHC-Based QoCS Strategy}\label{subsec:qocs}
Each publisher CAV $i$ utilizes hotbooting PHC to search its optimal QoCS strategy on each shared content $c \in {{\cal C}_i}$ in the dynamic game through trial and error.
The state vector $\tilde{\bf{z}}_{i}^t = \left[\tilde{\bf{z}}_{i,1}^t,\cdots,\tilde{\bf{z}}_{i,c}^t,\cdots,\tilde{\bf{z}}_{i,C_i}^t \right]$ for publisher CAV $i$ consists of the previous payment sequences of subscriber CAV groups, i.e., $\tilde{\bf{z}}_{i,c}^t = {\bf{p}}_{i,c}^{t-1}$.
Here, $\tilde{\bf{z}}_{i,c}^t = \left[({\tilde{z}_{i,c}^1})^t,({\tilde{z}_{i,c}^2})^t \right]$, and $({\tilde{z}_{i,c}^u})^t = ({{p}_{i,c}^u})^{t-1}$, $\forall u \in \{1,2\}$.
For simplicity, the QoCS strategy of each publisher CAV $i$ is uniformly quantized into $Y+1$ levels, i.e., ${{q}_{i,c}^u} \in \mathcal{Y}=\{\frac{y}{Y} \}_{0 \le y \le Y}$, $\forall u \in \{1,2\}$. The reward of publisher CAV $i$ is defined as its scaled utility $\lambda_2 {\mathcal{U}_i}\left({{\bf{q}}_i^t}, {{\bf{p}}_{i}^t} \right)$, where ${{\bf{q}}_{i}^t} = \left[{\mathbf{q}_{i,1}^t}, \cdots, {\mathbf{q}_{i,c}^t}, \cdots, {\mathbf{q}_{i,{C}_i}^t}\right]$ and $\lambda_2$ is a positive scale factor.

Let $\tilde{\mathbb{{Q}}}\left(\tilde{\bf{z}}_{i,c}^t, {\bf{q}}_{i,c}^t\right)$ denote the Q-function of state $\tilde{\bf{z}}_{i,c}^t$ and action ${\bf{q}}_{i,c}^t = \left[({{q}_{i,c}^1})^t,({{q}_{i,c}^2})^t \right]$. According to iterative Bellman equation, the Q-function can be updated as follows:
\begin{align}\label{eq:6-5}
\tilde{\mathbb{{Q}}}\left(\tilde{\bf{z}}_{i,c}^t, {\bf{q}}_{i,c}^t\right) &\leftarrow \tilde{\mathbb{{Q}}}\left(\tilde{\bf{z}}_{i,c}^t, {\bf{q}}_{i,c}^t\right) + \psi_2 \left\{ \lambda_2{\mathcal{U}_i}\left({{\bf{q}}_i^t}, {{\bf{p}}_{i}^t} \right) \right. \nonumber \\
&\left. {+ \chi_1 \tilde{\mathbb{{V}}} \left(\tilde{\bf{z}}_{i,c}^{t+1} \right) - \tilde{\mathbb{{Q}}}\left(\tilde{\bf{z}}_{i,c}^t, {\bf{q}}_{i,c}^t\right)} \right\}, \forall c \in {{\cal C}_i},
\end{align}
where $\psi_2 \in (0,1]$ is the learning rate, and $\chi_2 \in [0,1]$ is the discount factor. $\tilde{\bf{z}}_{i,c}^{t+1}$ is the new state vector of subscriber CAV group ${{\cal J}_c}$ at time slot $t+1$, which is transformed from state $\tilde{\bf{z}}_{i,c}^{t}$ with action ${\bf{q}}_{i,c}^t$.
The value function $\tilde{\mathbb{{V}}} \left(\tilde{\bf{z}}_{i,c}^{t+1} \right)$ which maximizes the Q-function at state $\tilde{\bf{z}}_{i,c}^t$ over the action set is updated by
\begin{align}\label{eq:6-6}
\tilde{\mathbb{{V}}} \left(\tilde{\bf{z}}_{i,c}^{t+1} \right) \leftarrow \mathop {\max }\limits_{{\bf{q}}_{i,c}} \tilde{\mathbb{{Q}}} \left( \tilde{\bf{z}}_{i,c}^{t+1}, {\bf{q}}_{i,c}^{t+1} \right).
\end{align}

Similarly, the mixed-strategy table $\tilde{\pi}\left(\tilde{\bf{z}}_{i,c}^t, {\bf{q}}_{i,c}^t\right)$ in PHC is updated by increasing the probability of behaving greedily by a small value $\delta_2$, $0 < \delta_2 \leq 1$, and decreasing the other probabilities by $-\frac{\delta_2}{Y}$. We have
\begin{align}\label{eq:6-7}
\tilde{\pi}&\left(\tilde{\bf{z}}_{i,c}^t, {\bf{q}}_{i,c}^t\right) \gets \tilde{\pi}\left(\tilde{\bf{z}}_{i,c}^t, {\bf{q}}_{i,c}^t\right)\nonumber \\
&+\left\{ \begin{array}{cl}
	\delta _2,&if\ \mathbf{q}_{i,c}^{*}=\arg\max_{\mathbf{q}_{i,c}}\tilde{\mathbb{{Q}}}\left(\tilde{\bf{z}}_{i,c}^t, {\mathbf{q}_{i,c}}\right);\\[0.02cm]
	-\frac{\delta _2}{Y+1},&otherwise.
\end{array} \right.
\end{align}
Then, the publisher CAV $i$ selects its QoCS strategy ${\bf{q}}_{i,c}^t$ on content $c$ based on the mixed-strategy table $\tilde{\pi}\left(\tilde{\bf{z}}_{i,c}^t, {\bf{q}}_{i,c}^t\right)$, i.e.,
\begin{align}\label{eq:6-8}
\Pr\left( \mathbf{q}_{i,c}^{t} = \hat{\bf{q}}_{i,c} \right) = \pi \left( \tilde{{\bf{z}}}_{i,c}^t,\hat{\bf{q}}_{i,c} \right) ,\forall \hat{\mathbf{q}}_{i,c}\in \mathcal{Y}.
\end{align}

A hotbooting technique (as shown in lines 14--20 in Algorithm 1) is also utilized by publisher CAVs to initialize the Q-value and mixed-strategy table to speed up the learning process by exploiting the historical experience.
The output of Algorithm 1 through $W$ experiments (i.e., $\tilde{\mathbb{{Q}}}_h$ and $\tilde{\pi}_h$) is utilized as the input of Algorithm 2, with $\tilde{\mathbb{{Q}}} = \tilde{\mathbb{{Q}}}_h$ and $\tilde{\pi} = \tilde{\pi}_h$. The two-tier hotbooting PHC-based optimal strategy decision algorithm for both publisher CAVs and subscriber CAVs is summarized in Algorithm \ref{Algorithm2}, where lines 14--21 shows the hotbooting PHC-based optimal QoCS strategy decision {process} for publisher CAV $i$.

\section{PERFORMANCE EVALUATION}\label{sec:SIMULATION}
In this section, we carry out extensive simulations to evaluate the performance of SPAD by using Matlab. The simulation setup is first introduced, followed by the numerical results and discussions.

\subsection{Simulation Setup}\label{subsec:evalution1}
\begin{table}[!t]\begin{small}\setlength{\abovecaptionskip}{-0.15cm}
    \begin{center}
        \caption{Simulation Parameters}\label{table1}
        \begin{tabular}{c c||c c}
        \hline \hline
        \textbf{Parameter} & \textbf{Value}  & \textbf{Parameter} & \textbf{Value}  \\ \hline
            $ {\tau(c)} $&$1$ &$\kappa$ &$0.9$  \\
            $ X, Y, W$ &$16, 10, 5$ &$\delta _1, \delta _2$ &$0.01$  \\
            $ {sc_{i,\pi_c}} $&$[0,1]$ &$ pc_i$&$[0,1]$   \\
            $ {\lambda _R} $&$0.05$ &$ {\lambda _B}$&$0.5$   \\
            $ {\eta _1},{\eta _2} $&$0.001$ &$ {w_1}, {w_2}, {w_3}$&$1$   \\
            $ \gamma $&$1.2$ &$ \theta_{\mathcal{J}_c}$&$0.45$  \\
            $ {\vartheta _1},{\vartheta _2} $&$0.75$ &$ {\gamma_1}, {\gamma_2}$&$0.01$   \\
            $ {\varepsilon _{i,\pi_c}^1},{\varepsilon _{i}^2} $&$[0.4,2.0]$ cents &$ {v}_a $&$[1,10]$  \\
            $ {\alpha _{i,c}} $&$[25,45]$ &$ {{\xi_c^1}},{{\xi_c^2}}$&$1$  \\
            $ s_c^1 $&$[0.1,0.5]$ MBytes &$ s_c^2 $&$[1,20]$ KBytes \\ 
            $ B_i^{Tr} $&$2$ MHz \cite{7968385} &{$\Im,\mathrm{SINR}_\lambda$}&{$4,100$ \cite{7968385}} \\ 
            $ P_i^{Tr} $&$23$ dBm \cite{Park2019simu4} &$ \sigma^2$&$-110$ dBm \cite{Park2019simu4} \\
            {$ a_n $}&{$0$ $\mathrm{m}/\mathrm{s}^2$} &{$ L_f,L_r$}& {$1.105,1.738$ m}  \\
            $ {\phi _0} $&$0.1$ cents &$ p_{\max}$&$5$ cents  \\
            $ \tilde{p}_1,\tilde{p}_2 $&$1.2$ cents &$ \beta_{j,c}$&random in $\{0,1\}$  \\
            \hline \hline
        \end{tabular}
    \end{center}\end{small}\vspace{-4mm}
\end{table}

We consider a simulation scenario with $100$ road segments in an actual urban area of San Francisco with about $11.03\times 7.06$ $km^2$ \cite{Traceset2009}. 
The length of each segment follows the uniform distribution and lies in $[20,200]$m. MEC nodes are evenly deployed along road segments every $200$m and have the same coverage radius $100$m \cite{Su2018}. The vehicle density of each road segment is randomly picked within $[10,120]$ veh/km. 
All CAVs in a CAV fleet drive at the constant velocity. 
The minimum and maximum velocities of a CAV fleet are set as $50$ and $110$ km/h, respectively.
As referenced in 3GPP LTE-V2X standard \cite{Park2019simu4}, the transmission power and the noise power are set as $23$ dBm and $-110$ dBm, respectively. Each CAV publishes a raw sensory content with its processed results in its fleet per time slot.
Each CAV randomly selects to subscribe to a published content or not in its current fleet. All subscriber CAVs of each published content randomly subscribe to its raw sensing part or the processed results part.
Based on \cite{8647913}, the large Zipf parameter (i.e., $\kappa =0.9$) is adopted for better content caching performance. 
The punishment factor and reputation threshold are set as $1.2$ and $0.45$, respectively. The cost parameter of each publisher CAV follows the uniform distribution ranged from $0.4$ to $2.0$ cents.
In the hotbooting PHC, the learning rates are set as $\psi_1=\psi_2=0.7$, and the discount factors are set as $\chi_1=\chi_2=0.7$.
Parameters in the simulation are summarized in Table \ref{table1}.

In the simulation, we consider three types of CAVs with different security levels, i.e., legitimate CAVs, speculative CAVs, and malicious CAVs. Legitimate CAVs always provide secure and true content for subscriber CAVs, while malicious ones may conduct attacks by delivering false or harmful content. Speculative CAV randomly opts to be legitimate or malicious. Let $r_l$, $r_s$, and $r_m$ be the ratios of legitimate, speculative, and malicious CAVs, respectively.
The following security metrics are utilized to indicate the performance of SPAD in defending against attacks defined in Sect. \ref{subsec:securitymodel}.
\begin{itemize}
  \item To evaluate the dependability of shared content, the following metric is employed. \emph{Secure pub/sub ratio: }the proportion of successfully {published/subscribed} secure and true content to the total number of {published/subscribed} contents among CAVs. The performance of SPAD in defending the \emph{dishonest and harmful content publishing attack} can refer to Figs. \ref{fig:simu1} and \ref{fig:simu2} in the next subsection. 
  \item High-quality {published} data can help subscriber AVs attain high-accurate information about their driving environment. As analyzed in Sect. \ref{subsec:contentmodel}, the notation ${\mathbf{q}_{i,c}}$ is utilized to measure the \emph{quality of content service (QoCS)} of publisher CAV $i$ in contributing the raw sensory data $c$ and its processed results. The performance of SPAD in defending the \emph{meaningless and low-quality content publishing attack} can refer to Figs. \ref{fig:simu4} and \ref{fig:simu8} in the next subsection.
\end{itemize}

The performance of the proposed SPAD scheme is evaluated by comparing with the following conventional schemes:
\begin{itemize}
  \item \textit{Traditional Bayesian inference-based trust (BIT) scheme \cite{8358773}.} In BIT scheme, the trustworthiness of each vehicle is predicted based on its historical behaviors by using the well-founded {standard} Bayesian inference mechanism, as shown in Eqs. (\ref{eq:beta})--(\ref{eq:bayes-e}), whereas vehicle's social role effects, time fading effects, and punishment for misbehaviors are not taken into account. Besides, the proposed static game $\mathbb{G}$ is applied in BIT scheme for {publishing/subscribing} of CAVs. Here, ${\lambda _R}=0$, $ {\lambda _B}=1$, ${w_2}=0$, and other parameters keep unchanged. 
  \item \textit{Stackelberg game-based scheme without reputation evaluation (SWR).} In SWR scheme, the interaction between publisher CAV and subscriber CAVs is modeled by the two-stage Stackelberg game $\mathbb{G}$ while the reputation assessment is not considered. 
  \item \textit{Two-tier Q-learning scheme \cite{9035635}.} In this scheme, the traditional Q-learning with $\epsilon$--greedy policy is employed for both publisher CAVs and subscriber CAVs to seek their optimal strategies in the dynamic game. Here, both the learning rate and discount factor remain unchanged. 
  \item \textit{Greedy scheme.} In this scheme, both publisher CAVs and subscriber CAVs behave greedily to seek their optimal strategies in the dynamic game based on Q-learning.
  \item \textit{Fixed price (FP) scheme.} In FP scheme, all subscriber CAVs pay content contributors with fixed content price vector ${\tilde {\mathbf{p}}} = [{\tilde {p_1}},\,{\tilde {p_2}}]$ in the static game, where ${\tilde {p_1}},{\tilde {p_2}}$ are the fixed prices of $raw_c$ and $result_c$ of content $c$, respectively.
\end{itemize}

\begin{table}
    \centering
    \caption{{Computation and communication overheads of SPAD}}\label{tableoverhead}
{\begin{tabular}{|c|c|c|c|}
\hline
& \textbf{\begin{tabular}[c]{@{}c@{}}Reputation \\ mechanism\end{tabular}} & \textbf{\begin{tabular}[c]{@{}c@{}}Static game \\ model\end{tabular}} & \textbf{Dynamic game model}                                              \\ \hline
\textbf{\begin{tabular}[c]{@{}c@{}}Computation \\ complexity\end{tabular}} & $\mathcal{O}(N)$                                                                     & $\mathcal{O}(C)$                                                                  & $\mathcal{O}(C\cdot T)$                                                                    \\ \hline
\textbf{Execution time}                                                      & 28 ms                                                                    & 11 ms                                                                 & \begin{tabular}[c]{@{}c@{}}135 ms (Hotbooting\\ with $W$\,=\,400)\end{tabular} \\ \hline
\textbf{\begin{tabular}[c]{@{}c@{}}Communication \\ complexity\end{tabular}} & $\mathcal{O}(N)$                                                                     & $\mathcal{O}(C)$                                                                  & $\mathcal{O}(C\cdot T)$                                                                    \\ \hline
\textbf{\begin{tabular}[c]{@{}c@{}}Communication \\ cost\end{tabular}}       & 20.5 MB                                                                  & 32 B                                                                  & \begin{tabular}[c]{@{}c@{}}384 B (Hotbooting\\ with $W$\,=\,400)\end{tabular}                                                                   \\ \hline
\end{tabular}}
\end{table}

\subsection{Numerical Results}\label{subsec:evalution2}
\begin{figure*}[!t]\setlength{\abovecaptionskip}{-0.0cm}
\begin{minipage}[!t]{0.32\textwidth}
\centering
    \includegraphics[height=4.2cm,width=\textwidth]{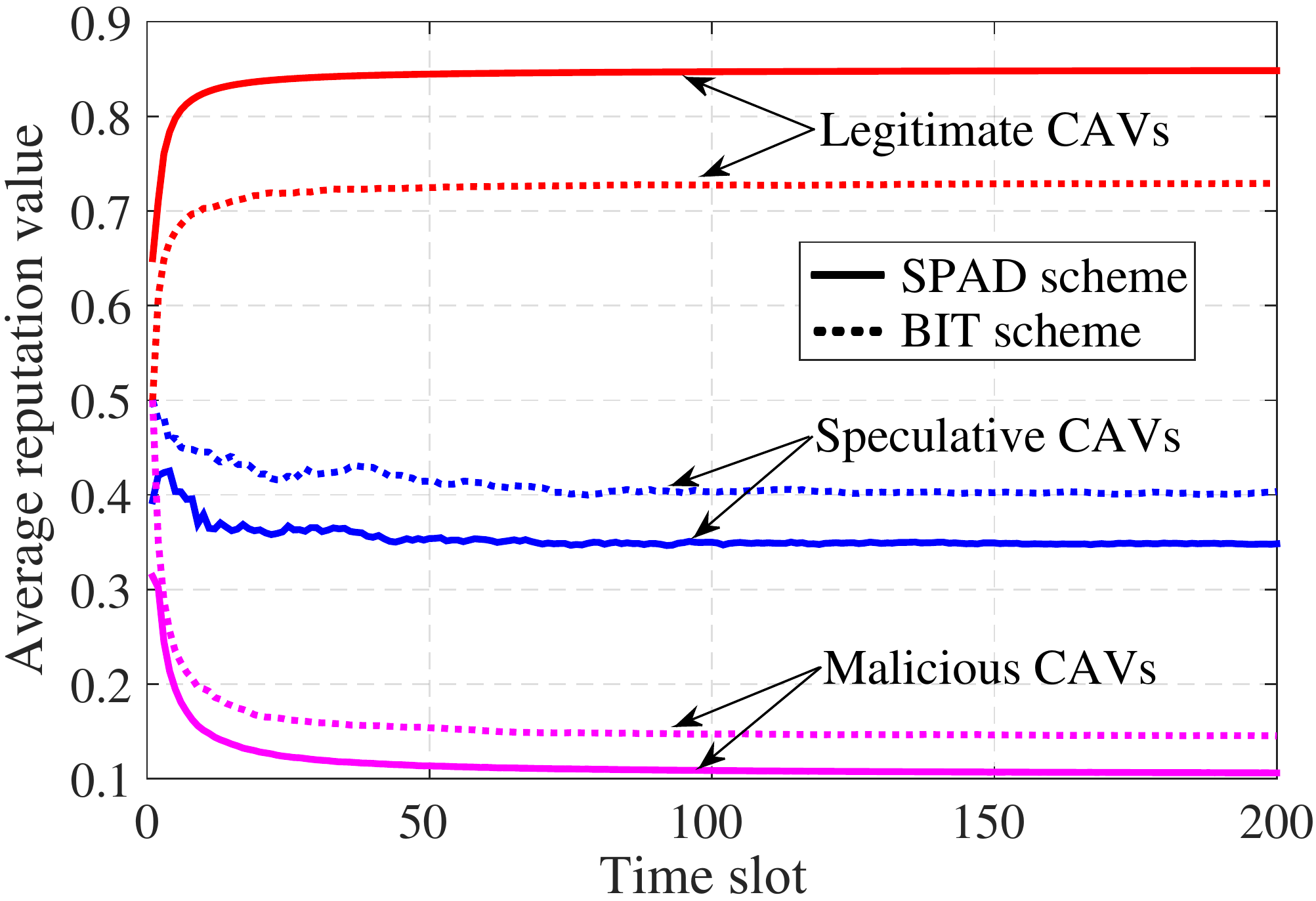} 
    \caption{Evolution of average reputation value over time for three types of CAVs, compared with {the} BIT scheme.}\label{fig:simu1}
\end{minipage}~~~~
\begin{minipage}[!t]{0.325\textwidth}
\centering
    \includegraphics[height=4.2cm,width=\textwidth]{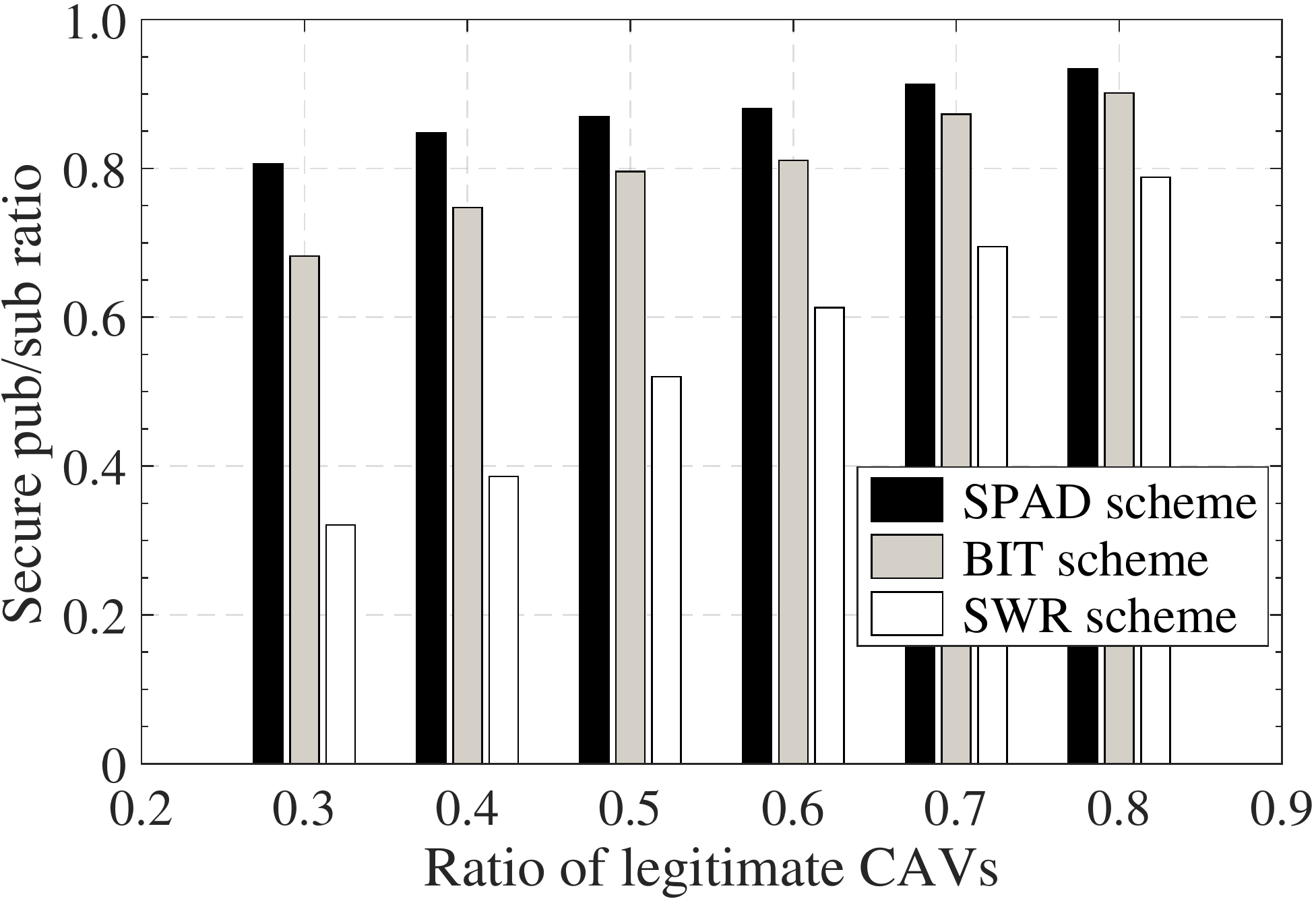}
    \caption{Secure pub/sub ratio vs. ratio of legitimate CAVs, compared with two existing schemes.}\label{fig:simu2}
\end{minipage}~~~~
\begin{minipage}[!t]{0.32\textwidth}
\centering
    \includegraphics[height=4.2cm,width=\textwidth]{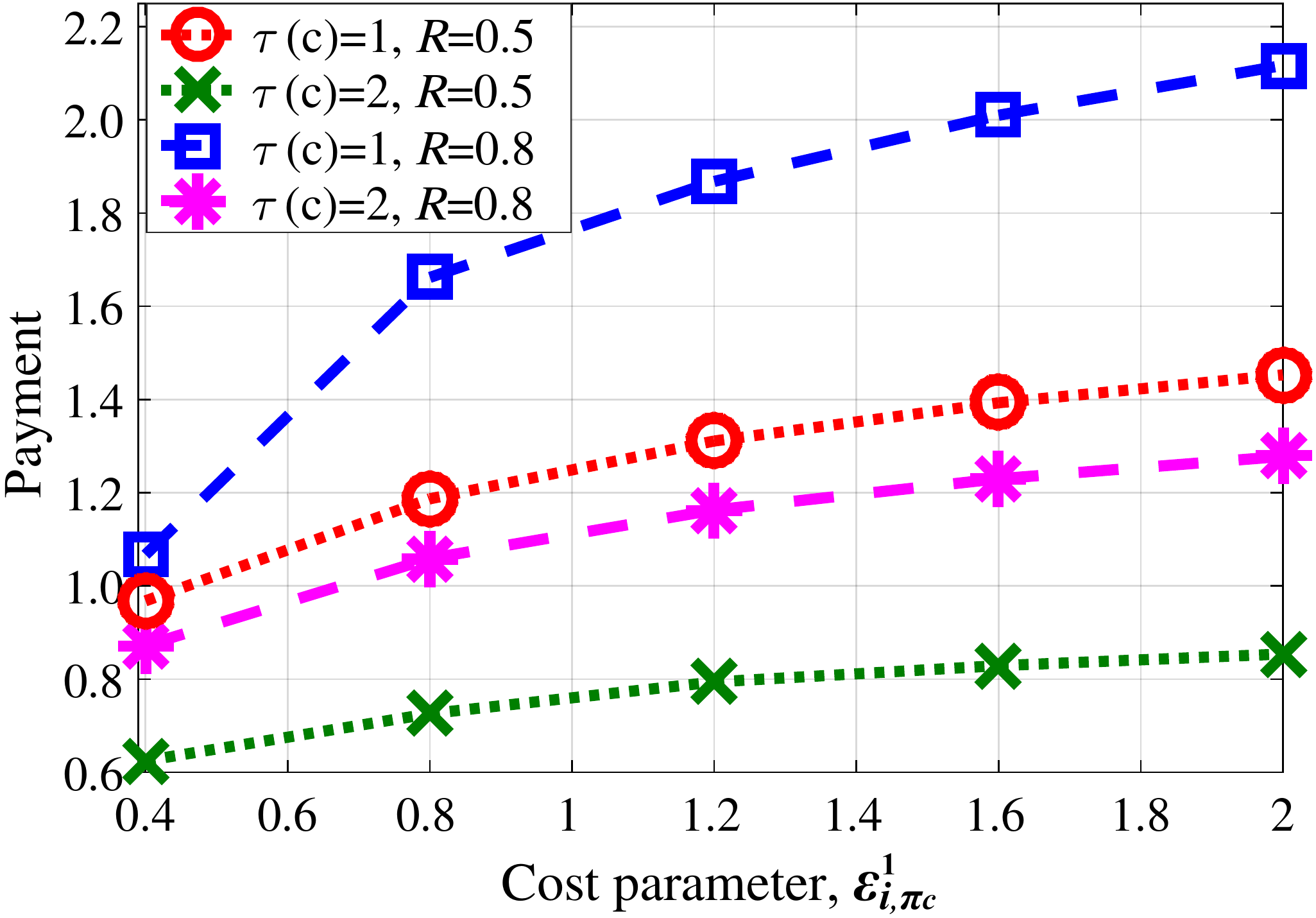}
    \caption{Optimal payment strategy of subscriber CAV vs. cost parameter of publisher CAV in static vehicular pub/sub game.}\label{fig:simu3}
\end{minipage}
\end{figure*}

{Table~\ref{tableoverhead} shows computation and communication overheads of our SPAD scheme in terms of reputation model, static pub/sub game model, and PHC-based dynamic pub/sub game model. For the reputation model, the computation and communication complexities at each time slot yield to $\mathcal{O}(N)$, where $N$ is the total number of CAVs. Besides, the reputation of CAVs can be evaluated and updated in parallel, thereby further reducing the running time. For the static game model, both publisher CAV and subscriber CAV calculate their optimal strategies by obeying the SE in a distributed manner. Both the computation and communication complexities for each CAV yield to $\mathcal{O}(C)$, where $C$ is the number of contents to be published/subscribed. Besides, for each publisher/subscriber CAV, it can compute the optimal strategies for different contents in parallel, thereby further reducing the execution time. For the PHC-based dynamic pub/sub game model, both publisher CAV and subscriber CAV calculate their near-optimal strategies by employing the policy hill-climbing (PHC) in a distributed manner. Similarly, both the computation and communication complexities for each CAV yield to $\mathcal{O}(C\cdot T)$, where $T$ is the number of convergent time slots. Here, the hotbooting technique is employed to speed up the convergence time by learning from historical experience. 
Thereby, both the running time and signaling overhead can be reduced. As seen in Table~\ref{tableoverhead}, both the communication and communication overheads for CAVs are very small. Thereby, the consumed bandwidth and computation resources of CAVs by employing our SPAD scheme can be very small, which validates the practicality of our SPAD scheme.}

Fig. \ref{fig:simu1} depicts the evolution of average reputation values over time for three types of CAVs in two schemes. Here, we set $r_l \!=\! 0.6$, $r_s \!=\! 0.2$, and $r_m \!=\! 0.2$.
As seen in Fig. \ref{fig:simu1}, in both two schemes, the average reputation values of speculative CAVs and malicious CAVs decline over time, while that of legitimate CAVs ascents. The reason is that legitimate CAVs always behave honestly and cooperatively to gradually increase their reputation values. In opposite, the misbehaviors conducted by malicious CAVs will be recorded for reputation calculation, resulting in low reputation values of malicious CAVs. Moreover, the average reputation value of speculative CAVs is relatively higher than that of malicious CAVs, since speculative CAVs can randomly choose their actions to be legitimate or malicious.

Furthermore, as seen in Fig. \ref{fig:simu1}, in the proposed SPAD scheme, the average reputation value of legitimate CAVs is larger than that in {the} BIT scheme, while the average reputation values of speculative or malicious CAVs {converge faster and are} smaller than that in {the} BIT scheme. 
{Moreover, in our} SPAD scheme, the initial average reputation value of legitimate CAVs (i.e., $0.6496$) is higher than that of speculative CAVs (i.e., $0.3937$) and malicious CAVs (i.e., $0.3137$); while in {the} BIT scheme, the initial average reputation values of three types of CAVs are identical (i.e., $0.5$). 
{It can be explained as follows.
Firstly, in our SPAD scheme, CAVs are featured with diversified social roles $\mathcal{A}$. {Consequently,} legitimate CAVs are assigned to social role types with higher trustworthiness degree ${v}_a$, while speculative and malicious CAVs correspond to those with smaller trustworthiness degrees.
Secondly, as SPAD additionally considers time fading effects in reputation evaluation, it leads to the fading weights of historical behaviors in determining the current reputation value and thereby faster convergence speed for speculative CAVs' reputation.
Thirdly, a punishment factor is further incorporated in reputation evaluation, causing a larger reputation drop for both speculative and malicious CAVs when their misbehaviors are detected and higher difficulty in recovering their reputation by behaving honestly in following time slots. As a consequence, the average reputation values of both speculative and malicious CAVs in our SPAD scheme are smaller than that in {the} BIT scheme.
To summarize, the proposed SPAD scheme can attain improved accuracy and robustness in reputation evaluation.}

Fig. \ref{fig:simu2} illustrates the comparison of SPAD scheme with the other two {baselines} on the secure {pub/sub} ratio, where the ratio of legitimate CAVs varies from $0.4$ to $0.8$. In this simulation, $r_s$ is fixed and equals to $0.2$. Other settings keep unchanged. We can see that the proposed SPAD scheme attains a higher {secure pub/sub ratio} than {both BIT and SWR schemes}, given different values of $r_l$. The reason is that in {the} BIT scheme, without considering both the time decay effect and punishment factor for misbehaving CAVs, malicious and speculative publisher CAVs may {contribute false data} in older time slots and quickly recover their reputation values by behaving honestly in recent time. Meanwhile, the dimension of {vehicle's social role} is not considered in reputation computing in {the} BIT scheme, which may result in a degradation of accuracy in reputation evaluation results.
In {the} SWR scheme, due to the absence of reputation assessment, the selected publisher CAVs may deliver false content to cheat subscriber CAVs, resulting in a low secure {pub/sub} ratio.
{In opposite, our SPAD scheme builds a hybrid reputation model by considering the effects of vehicle's social roles, behaviors, time fading, and misbehavior punishment during reputation evaluation process, resulting in a more accurate reputation assessment result for different types of CAVs. Thereby, the dependability of disseminated vehicular contents can be improved in SPAD even if the ratio of legitimate CAVs is low.}

\begin{figure*}[htbp]\setlength{\abovecaptionskip}{-0.0cm}
\begin{minipage}[t]{0.32\textwidth}
\centering
    \includegraphics[height=4.2cm,width=\textwidth]{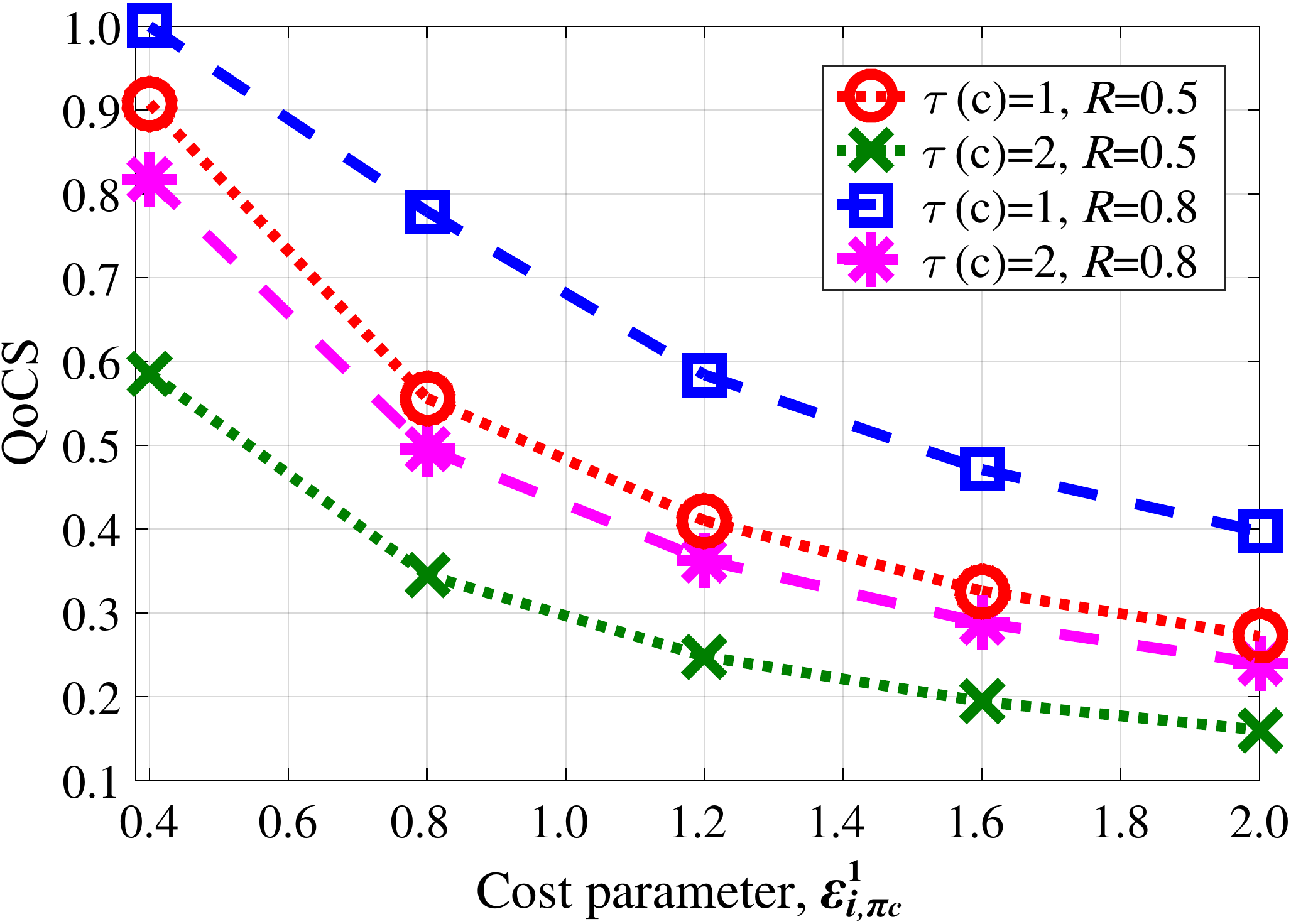}
    \caption{Optimal QoCS strategy of publisher CAV vs. cost parameter of publisher CAV in static vehicular pub/sub game.}\label{fig:simu4}
\end{minipage}~~~~
\begin{minipage}[t]{0.318\textwidth}
\centering
    \includegraphics[height=4.2cm,width=\textwidth]{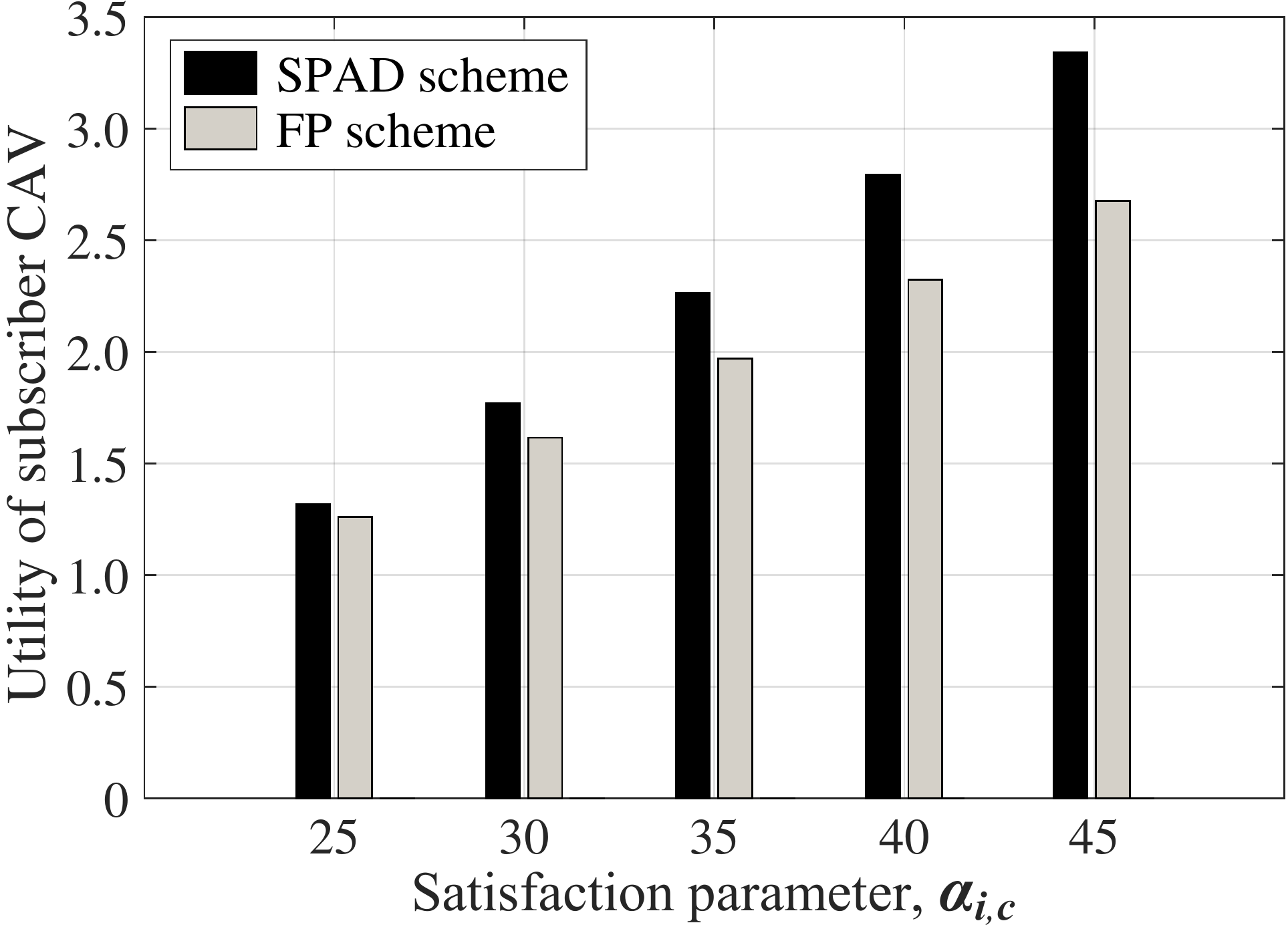}
    \caption{Utility of subscriber CAV vs. satisfaction parameter of subscriber CAV in static vehicular pub/sub game, compared with {the FP scheme}.}\label{fig:simu5}
\end{minipage}~~~~~
\begin{minipage}[t]{0.32\textwidth}
\centering
    \includegraphics[height=4.2cm,width=\textwidth]{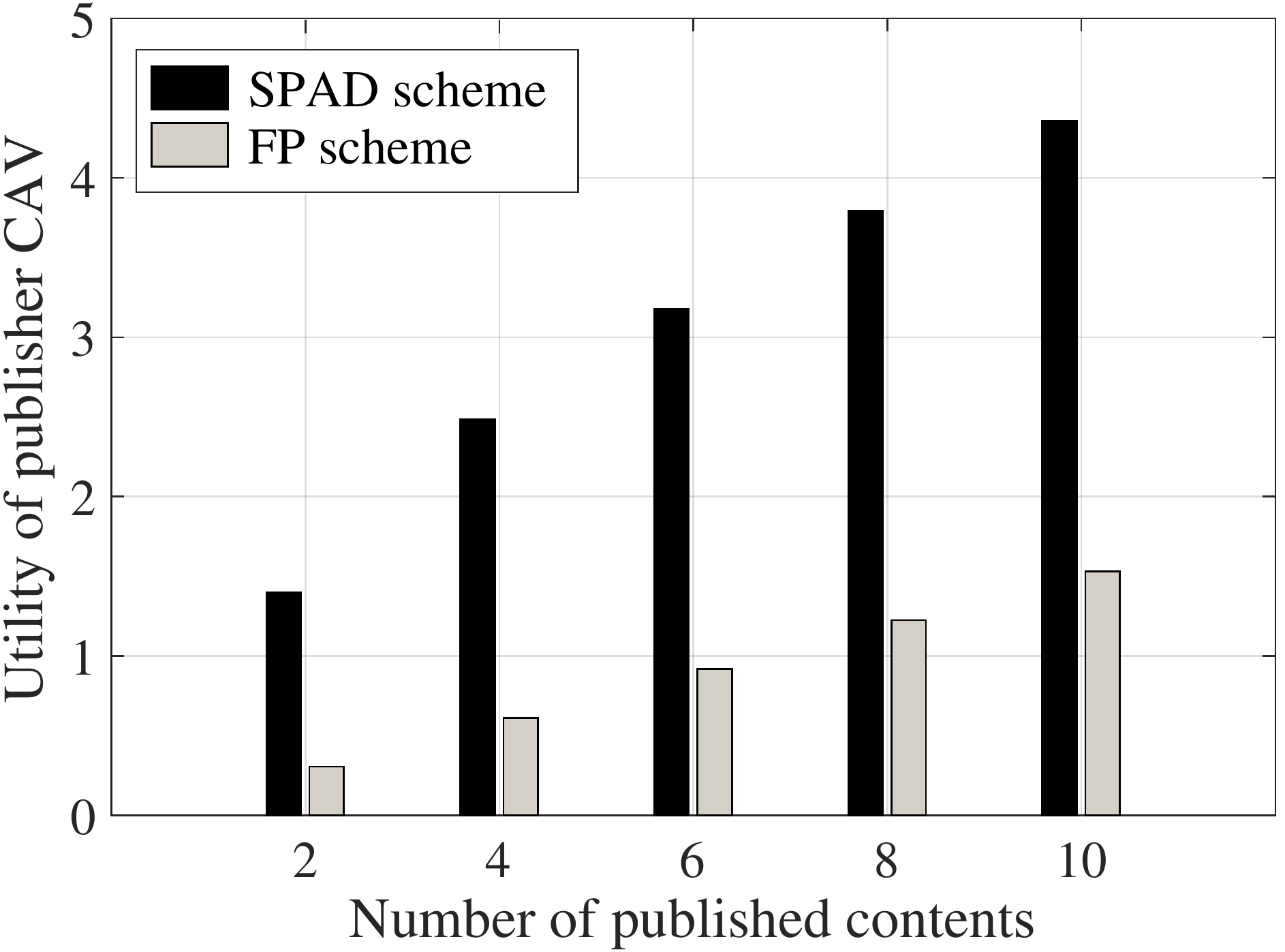}
    \caption{Utility of publisher CAV vs. number of published contents in static vehicular pub/sub game, compared with {the FP scheme}.}\label{fig:simu6}
\end{minipage}
\end{figure*}

\begin{figure*}[!tbp]\setlength{\abovecaptionskip}{-0.0cm}
\begin{minipage}[t]{0.32\textwidth}
\centering
    \includegraphics[height=4.1cm,width=\textwidth]{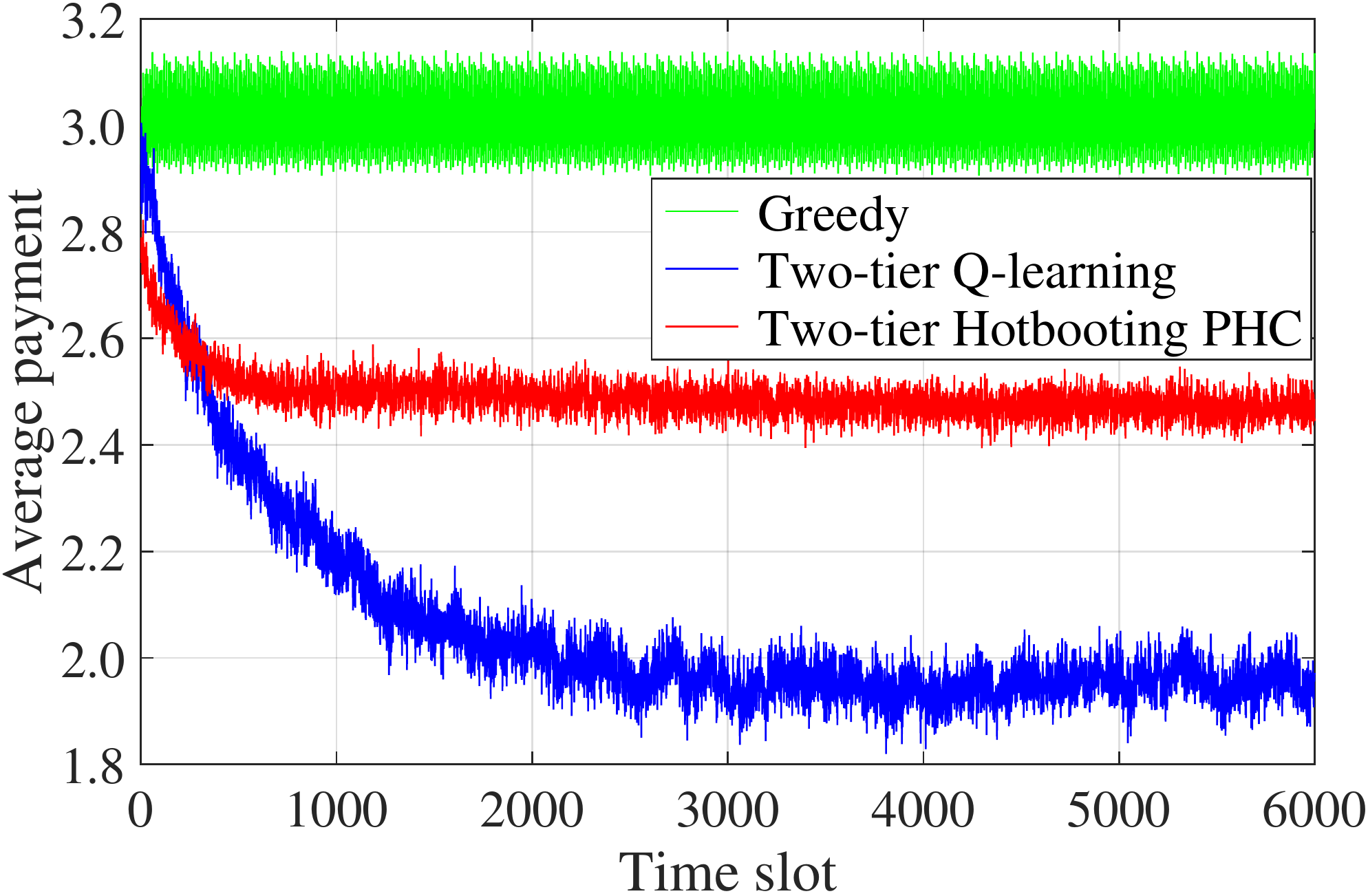}
    \caption{Evolution of average payment of subscriber CAVs in dynamic game, compared with two existing schemes.}\label{fig:simu7}
\end{minipage}~~~~
\begin{minipage}[t]{0.318\textwidth}
\centering
    \includegraphics[height=4.1cm,width=\textwidth]{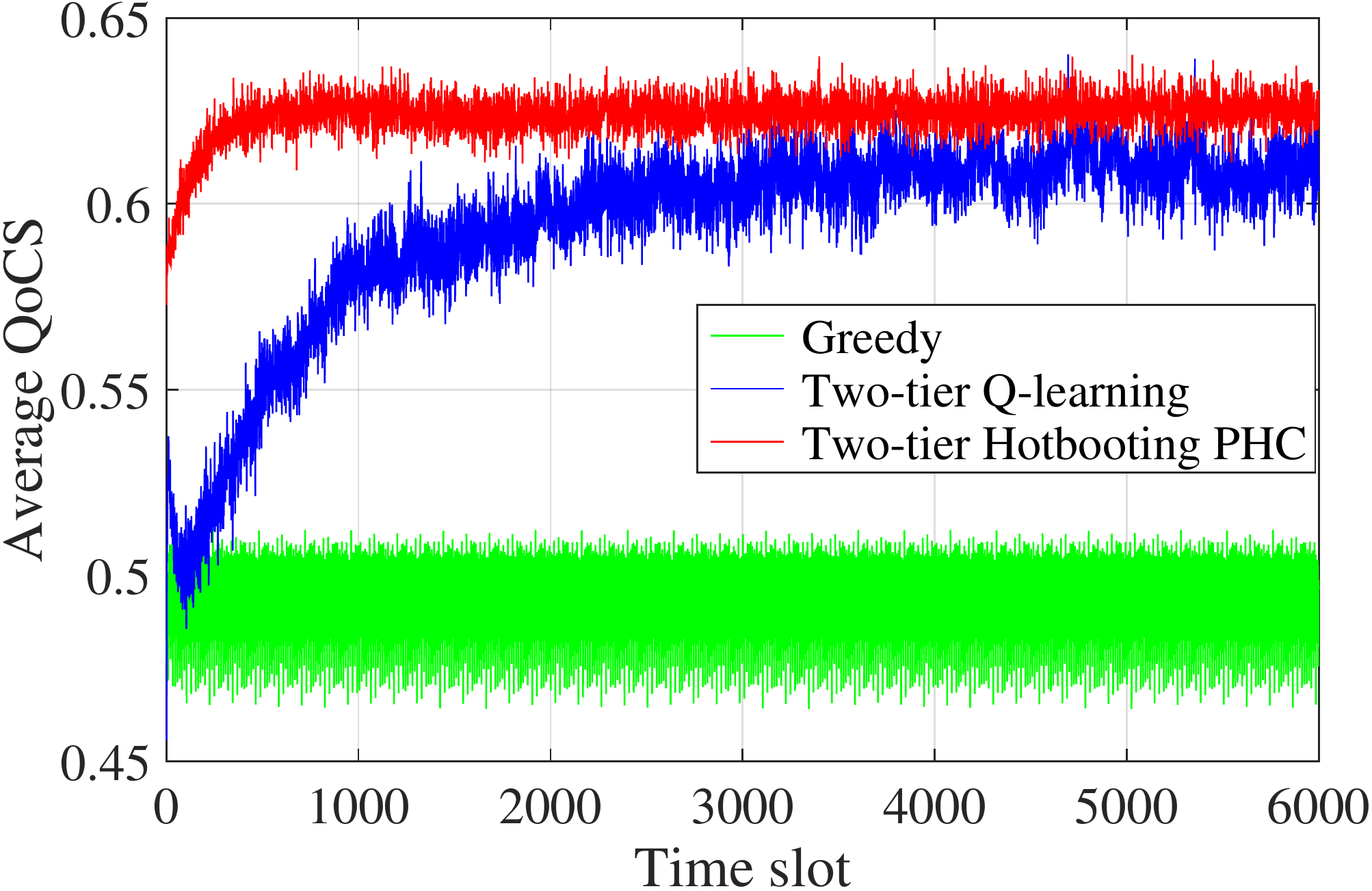}
    \caption{Evolution of average QoCS of publisher CAVs in dynamic game, compared with two existing schemes.}\label{fig:simu8}
\end{minipage}~~~~~
\begin{minipage}[t]{0.32\textwidth}
\centering
    \includegraphics[height=4.1cm,width=\textwidth]{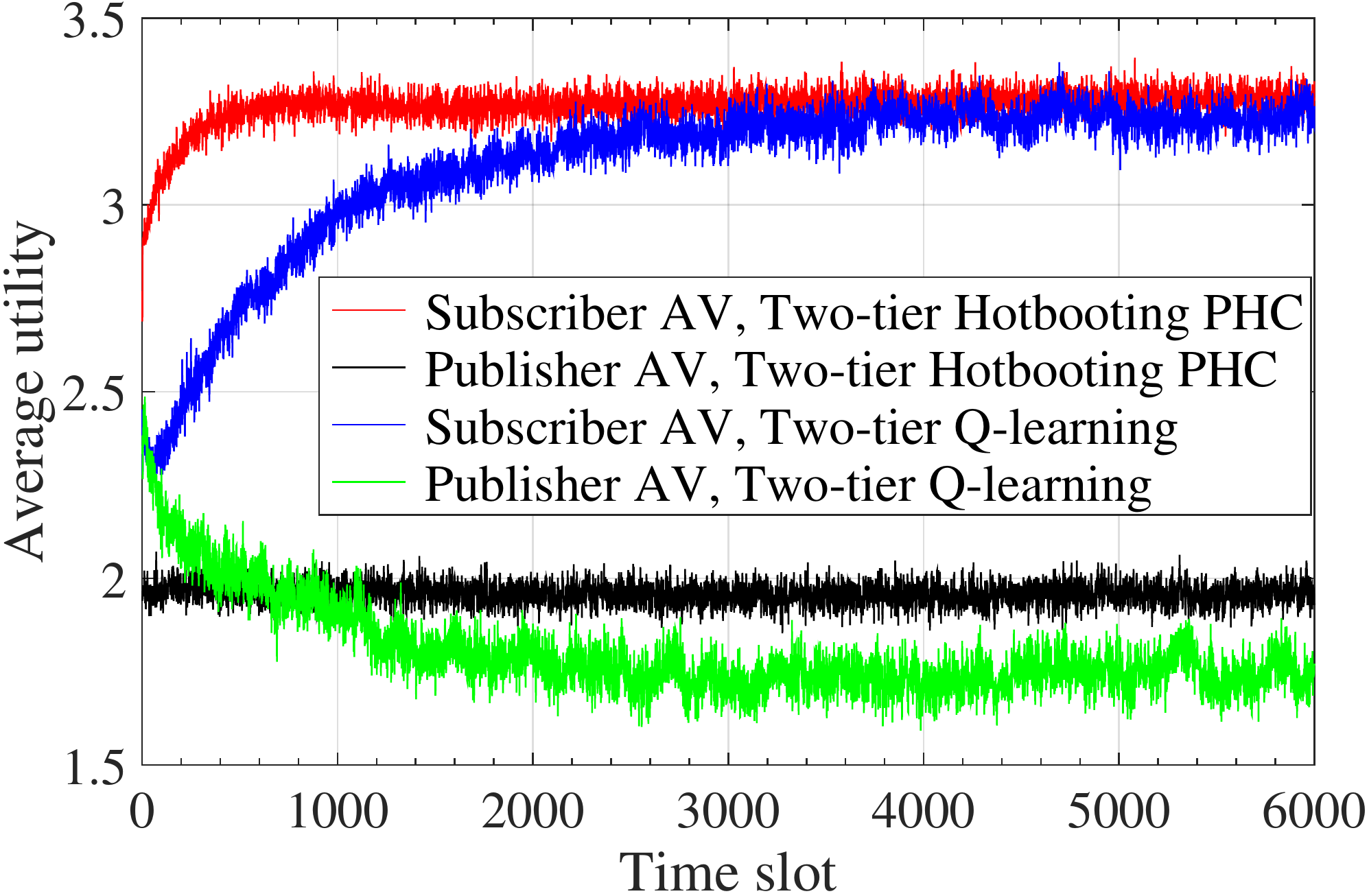}
    \caption{Evolution of average utility of pub/sub CAVs in dynamic game, compared with two-tier Q-learning scheme.}\label{fig:simu9}
\end{minipage}
\end{figure*}

Fig. \ref{fig:simu3} and Fig. \ref{fig:simu4} demonstrate the optimal payment strategy of subscriber CAV and optimal QoCS strategy of publisher CAV on raw sensory data $raw_c$ in the static vehicular pub/sub game, respectively, where the cost parameter of publisher CAV changes from $0.4$ to $2.0$. Here, the satisfaction coefficient is set as ${\alpha _{i,c}} = 28$. Other settings are unchanged. From Figs. \ref{fig:simu3} and \ref{fig:simu4}, it can be seen that with the increase of cost parameter $\varepsilon_{i,\pi_c}^1$, the optimal payment of subscriber CAV increases, while publisher CAV's optimal QoCS is in a decline. The reason is that, since a higher cost parameter indicates a higher cost in contributing content, {each} publisher CAV prefers to decrease the QoCS to reduce its cost, while subscriber CAVs have a high willingness to pay more to stimulate high-quality content services. Besides, as smaller content index ${\tau(c)}$ means higher content popularity and higher reputation value $R$ implies higher trustworthiness, both the payment and QoCS in pub/sub services attain a higher value given a smaller ${\tau(c)}$ and a higher $R$, i.e., ${\tau(c)} = 1$ and $R = 0.8$. {The above observed results are in accord with the SE of the vehicular pub/sub game obtained in Eqs. (\ref{eq:q*1})--(\ref{eq:stackSE}).}


Fig. \ref{fig:simu5} shows the comparison of SPAD scheme with {the FP scheme} on the utility of subscriber CAV in the static game, where the satisfaction parameter of publisher CAV changes from $25$ to $45$.
Fig. \ref{fig:simu6} compares SPAD scheme with {the} FP scheme on the utility of publisher CAV, where the number of published contents of publisher CAV varies from $2$ to $10$. Here, ${\tau(c)}=1$ and $R=0.8$. Other settings keep unchanged. From Figs. \ref{fig:simu5} and \ref{fig:simu6}, it can be observed that our SPAD scheme outperforms {the} FP scheme by attaining higher utilities for both subscriber CAV and publisher CAV. The reason is that, as the content price ${\tilde {\mathbf{p}}}$ is fixed and unalterable in {the} FP scheme, both subscriber CAV and publisher CAV can only achieve local optimal utilities given ${\tilde {\mathbf{p}}}$, but not the global ones with the optimal content price. 
In the proposed SPAD scheme, content consumers and content contributors can determine the optimal payment and optimal quality strategies during {publishing/subscribing process} to maximize their utilities by obeying the SE {in Eq. (\ref{eq:stackSE})}, respectively.

In the next simulations for dynamic games, we set ${\alpha _{i,c}} = 42$, $\varepsilon_{i,\pi_c}^1=0.4$, and other settings are unchanged.
Figs. \ref{fig:simu7} and \ref{fig:simu8} depict the evolutions of learning-based payment and QoCS strategy-making processes over time, {respectively, in comparison} with the two-tier Q-learning scheme and greedy scheme.
Fig. \ref{fig:simu9} illustrates the evolutions of average utilities of subscriber CAVs and publisher CAVs over time, compared with {the} two-tier Q-learning scheme.
As seen from the three figures, the proposed two-tier hotbooting PHC algorithm can converge quickly in searching the optimal policies for players in the dynamic game after about $800$ time slots, while the convergence time in the two-tier Q-learning scheme is about $3000$ time slots. Accordingly, our proposed scheme can attain a much-improved learning speed and faster convergence rate.
The reason is that, our hotbooting PHC utilizes both previous experiences learned in similar scenarios and the mixed-strategy policy update to save the initial exploration time and accelerate the learning speed.

Besides, as shown in Fig. \ref{fig:simu8}, the proposed scheme attains a slightly higher QoCS than {the} two-tier Q-learning scheme and a much-improved QoCS than {the} greedy scheme. It can be explained as follows. On one hand, due to the stable high payment after $800$ time slots (as shown in Fig. \ref{fig:simu7}), the publisher CAV is motivated to contribute contents with higher QoCS to seek maximized future benefits. Accordingly, the QoCS resulted from the dynamic game in Q-learning is slightly lower than our proposed scheme. On the other hand, as CAVs are behaving greedily to seek the maximized immediate reward and overlook future benefits, the QoCS (resp. payment) is much lower (resp. higher) than {the} other two schemes.
In addition, in Fig. \ref{fig:simu9}, we can observe that our proposed scheme can attain slightly higher utilities for both subscriber CAVs and publisher CAVs than {the} two-tier Q-learning scheme.
It can be explained as below. Due to the high payment and QoCS of delivered contents in our hotbooting PHC scheme (as seen in Figs. \ref{fig:simu7} and \ref{fig:simu8}), under current system parameter settings, both subscriber CAVs and publisher CAVs can attain improved utilities in the game than {the} two-tier Q-learning scheme.

\section{Conclusion and Future Work}\label{sec:CONSLUSION}
In this paper, we have proposed SPAD, a secured {cooperative autonomous driving framework}.
Firstly, to stimulate CAVs' participation and high-quality data publishing, the interactions between publisher CAVs and subscriber CAVs have been formulated as a vehicular pub/sub game. A reputation model has been also designed to punish CAVs' dishonest and harmful content publishing behaviors.
Secondly, the SE of the static vehicular pub/sub game has been analyzed to derive the optimal strategies and maximize the utilities of both publisher CAV and subscriber CAVs during content {publishing/subscribing}.
In addition, a two-tier hotbooting PHC algorithm has been devised to efficiently search the optimal strategies for each player in the dynamic game in practical networks, without the awareness of accurate network parameters and its opponents' utility model parameters.
Finally, simulation results have shown that the proposed SPAD scheme can achieve {free-rider prevention}, enhanced dependability for {subscribed contents}, improved {vehicles' utilities}, and faster learning speed, compared with conventional schemes.

For the future work, we will extend this work by considering the blockchain-based digital forensics for CAVs in reputation evaluation.
In addition, we plan to design the cooperation mechanism among CAV fleets, as well as MEC nodes, to further improve the security and efficiency in autonomous driving.


\begin{appendices}
\section{{proof of theorem $1$}}\label{Appendix A}
\begin{IEEEproof}
Note that the quality parameters ${q_{i,c}^u},\forall u \in \{1,2\}$ are decoupled. Here, we only consider the case that $u=1$, and the other case that $u=2$ can be proved similarly.
The first order differential for ${{\mathcal{U}_i}\left( {{{\bf{q}}_i},{{\bf{p}}_i}} \right)}$ with respect to ${q_{i,c}^1}$ is
\begin{align}\label{eq:}
\frac{{\partial {\mathcal{U}_i}\left( {{{\bf{q}}_i},{{\bf{p}}_i}} \right)}}{{\partial {q_{i,c}^1}}} = J_c^1{\vartheta_c^1} {p_{i,c}^1} - 2{\xi_c^1} {\varepsilon _{i,\pi_c}^1}{sc_{i,\pi_c}}{q_{i,c}^1}.
\end{align}
The second order differential for ${{\mathcal{U}_i}\left( {{{\bf{q}}_i},{{\bf{p}}_i}} \right)}$ with respect to ${q_{i,c}^1}$ satisfies
$\frac{{{\partial ^2}{\mathcal{U}_i}\left( {{{\bf{q}}_i},{{\bf{p}}_i}} \right)}}{{\partial {q_{i,c}^1}^2}} =  - 2{\xi_c^1} {\varepsilon _{i,\pi_c}^1}{sc_{i,\pi_c}} < 0$,
which implies that the utility function of publisher CAV $i$ is a strict convex function.
Furthermore, we have
\begin{numcases}{}
\mathop {\lim }\limits_{{q_{i,c}^1} \to 0} \frac{{\partial {\mathcal{U}_i}\left( {{{\bf{q}}_i},{{\bf{p}}_i}} \right)}}{{\partial {q_{i,c}^1}}} = J_c^1{\vartheta _c^1}{p_{i,c}^1} > 0,\\
\mathop {\lim }\limits_{{q_{i,c}^1} \to 1} \frac{{\partial {\mathcal{U}_i}\left( {{{\bf{q}}_i},{{\bf{p}}_i}} \right)}}{{\partial {q_{i,c}^1}}} = J_c^1{\vartheta _c^1}{p_{i,c}^1} - 2{\xi_c^1} {\varepsilon _{i,\pi_c}^1}{sc_{i,\pi_c}}.
\end{numcases}
Here, we consider the following two cases.

Case $1$: High payment. If the payment of subscriber CAV group ${{\mathcal{J}}_c^1}$ is high, i.e., ${p_{\max}} \ge {p_{i,c}^1} \ge \frac{2{\xi_c^1} {\varepsilon _{i,\pi_c}^1}{sc_{i,\pi_c}}}{{J_c^1{\vartheta_c^1} }}$, we have $\mathop {\lim }\limits_{{q_{i,c}^1} \to 1} \frac{{\partial {\mathcal{U}_i}\left( {{{\bf{q}}_i},{{\bf{p}}_i}} \right)}}{{\partial {q_{i,c}^1}}} \ge 0$. In this case, ${{\mathcal{U}_i}\left( {{{\bf{q}}_i},{{\bf{p}}_i}} \right)}$ is monotonically increasing with respect to ${q_{i,c}^1}$. Therefore, the optimal QoCS strategy of publisher CAV $i$ on $raw_c$ of content $c$ is ${q_{i,c}^1}^* = 1$.

Case $2$: Low payment. If the payment of subscriber CAV group ${{\mathcal{J}}_c^1}$ is low, i.e., $0 < {p_{i,c}^1} < \frac{2{\xi_c^1} {\varepsilon _{i,\pi_c}^1}{sc_{i,\pi_c}}}{{J_c^1{\vartheta_c^1} }}$, we have $\mathop {\lim }\limits_{{q_{i,c}^1} \to 1} \frac{{\partial {\mathcal{U}_i}\left( {{{\bf{q}}_i},{{\bf{p}}_i}} \right)}}{{\partial {q_{i,c}^1}}} < 0$. In this case, the optimal QoCS strategy of publisher CAV $i$ on content $c $ can be derived by solving $\frac{{\partial {\mathcal{U}_i}\left( {{{\bf{q}}_i},{{\bf{p}}_i}} \right)}}{{\partial {q_{i,c}^1}}} = 0$, i.e., 
\begin{align}\label{eq: responsefunc}
{q_{i,c}^1}^* =r\left({p_{i,c}^1} \right) = \frac{{J_c^1{\vartheta_c^1} {p_{i,c}^1}}}{2{\xi_c^1} {\varepsilon _{i,\pi_c}^1}{sc_{i,\pi_c}}},
\end{align}
where $r(.)$ is the optimal response function of publisher CAV $i$.
Theorem $1$ is proved.
\end{IEEEproof}

\section{{proof of theorem $2$}}\label{Appendix B}
\begin{IEEEproof}
Note that the price parameters ${p_{i,c}^u},,\forall u \in \{1,2\}$ are decoupled. We only consider the case when $u=1$, and the other case when $u=2$ can be proved in a similar manner.
Here, if ${p_{\max}} \ge {p_{i,c}^1} \ge \frac{2{\xi_c^1} {\varepsilon _{i,\pi_c}^1}{sc_{i,\pi_c}}}{{J_c^1{\vartheta_c^1} }}$, by substituting ${q_{i,c}^1}^* = 1$ into ${\mathcal{U}_{{{\cal J}_c}}}\left( {{\mathbf{q}_{i,c}}^*,{\mathbf{p}_{i,c}}} \right)$, the utility function of subscriber CAV group $\mathcal{J}_c$ can be rewritten as:
\begin{align}\label{eq:max2-1}
{\mathcal{U}_{{\mathcal{J}_c}}}&\left( {{{\bf{q}}_{i,c}}^*,{{\bf{p}}_{i,c}}} \right) = \nonumber \\
&{\alpha _{i,c}}{f_c}{R_i}\left({J_c^1}\log \left( {1 + s{c_{i,\pi_c}}}\right) + {J_c^2}\log \left( 1 + p{c_i}q{{_{i,c}^2}^*}\right) \right) \nonumber \\
& -{\left( {J_c^1\vartheta _c^1p_{i,c}^1 + J_c^2\vartheta _c^2p_{i,c}^2q{{_{i,c}^2}^*}} \right)}  -  \Theta, 
\end{align}
where
\begin{align}
\Theta = {\frac{1}{{r_{i,\mathcal{J}_c^1}^v}}{{J}_c^1\gamma ^1}s_c^1 } +  {\frac{1}{{r_{i,\mathcal{J}_c^2}^v}}{{J}_c^2\gamma ^2}s_c^2 } .
\end{align}
Since the above utility function is a monotonic decreasing function with respect to ${p_{i,c}^1}$, the optimal payment strategy of the subscriber CAV group $\mathcal{J}_c^1$ is denoted as ${p_{i,c}^1}^* = \frac{2{\xi_c^1} {\varepsilon _{i,\pi_c}^1}{sc_{i,\pi_c}}}{{J_c^1{\vartheta_c^1} }}$.

If $0 < {p_{i,c}^1} < \frac{2{\xi_c^1} {\varepsilon _{i,\pi_c}^1}{sc_{i,\pi_c}}}{{J_c^1{\vartheta_c^1} }}$, by substituting ${q_{i,c}^1}^* = \frac{{J_c^1{\vartheta_c^1} {p_{i,c}^1}}}{2{\xi_c^1} {\varepsilon _{i,\pi_c}^1}{sc_{i,\pi_c}}}$ into ${\mathcal{U}_{{{\cal J}_c}}}\left( {{\mathbf{q}_{i,c}}^*,{\mathbf{p}_{i,c}}} \right)$, the utility function of subscriber CAV group $\mathcal{J}_c$ can be reformulated as:
\begin{align}\label{eq:max2-2}
&{\mathcal{U}_{{\mathcal{J}_c}}}\left( {{{\bf{q}}_{i,c}}^*,{{\bf{p}}_{i,c}}} \right) = \nonumber \\
&{\alpha _{i,c}}{f_c}{R_i} \left({J_c^1}\log \left( {1 + \frac{{J_c^1{\vartheta_c^1} {p_{i,c}^1}}}{2{\xi_c^1} {\varepsilon _{i,\pi_c}^1}} } \right) + {J_c^2}\log \left( 1 + p{c_i}q{{_{i,c}^2}^*}\right) \right) \nonumber \\ 
& \!-\! {\left( {\frac{{{{\left( {J_c^1\vartheta _c^1p_{i,c}^1} \right)}^2}}}{{2{\xi_c^1} {\varepsilon _{i,\pi_c}^1}{sc_{i,\pi_c}}}} \!+\! J_c^2\vartheta _c^2p_{i,c}^2q{{_{i,c}^2}^*}} \right)} - \Theta.
\end{align}
The first order differential for ${\mathcal{U}_{{{\cal J}_c}}}\left( {{\mathbf{q}_{i,c}}^*,{\mathbf{p}_{i,c}}} \right)$ in Eq. (\ref{eq:max2-2}) with respect to ${p_{i,c}^1}$ is
\begin{align}\resizebox{.894\hsize}{!}{$
\frac{{\partial {\mathcal{U}_{{{\cal J}_c}}}\left( {{\mathbf{q}_{i,c}}^*,{\mathbf{p}_{i,c}}} \right)}}{{\partial {p_{i,c}^1}}} \!=\! \frac{{{\alpha _{i,c}}{f_c}{R_i}{\vartheta_c^1}{(J_c^1)^2} }}{{2{\xi_c^1} {\varepsilon _{i,\pi_c}^1} + {J_c^1}{\vartheta_c^1} {p_{i,c}^1}}} - \frac{{{(J_c^1\vartheta_c^1) ^2}{p_{i,c}^1}}}{{{\xi_c^1} {\varepsilon _{i,\pi_c}^1}{sc_{i,\pi_c}}}}. $}
\end{align}
The second order differential for ${\mathcal{U}_{{{\cal J}_c}}}\left( {{\mathbf{q}_{i,c}}^*,{\mathbf{p}_{i,c}}} \right)$ in Eq. (\ref{eq:max2-2}) with respect to ${p_{i,c}^1}$ satisfies
\begin{align}
\frac{{{\partial ^2}{\mathcal{U}_{{{\cal J}_c}}}\left( {{\mathbf{q}_{i,c}}^*,{\mathbf{p}_{i,c}}} \right)}}{{\partial {p_{i,c}^1}^2}} &\!=\! - \frac{{{\alpha _{i,c}}{f_c}{R_i}({\vartheta_c^1})^2{(J_c^1)^3} }}{\left({{2{\xi_c^1} {\varepsilon _{i,\pi_c}^1} \!+\! {J_c^1}{\vartheta_c^1} {p_{i,c}^1}}}\right)^2} \!-\! \frac{{{(J_c^1\vartheta_c^1) ^2}}}{{{\xi_c^1} {\varepsilon _{i,\pi_c}^1}{sc_{i,\pi_c}}}} \nonumber \\
&< 0,
\end{align}
which implies that the utility function of subscriber CAV group ${{\cal J}_c}$ in Eq. (\ref{eq:max2-2}) is strictly convex.
Furthermore, we have
\begin{numcases}{}
\mathop {\lim }\limits_{{p_{i,c}^1} \to 0} \frac{{\partial {\mathcal{U}_{{{\cal J}_c}}}\left( {{\mathbf{q}_{i,c}}^*,{\mathbf{p}_{i,c}}} \right)}}{{\partial {p_{i,c}^1}}} > 0, \\
\mathop {\lim }\limits_{{p_{i,c}^1} \to  + \infty } \frac{{\partial {\mathcal{U}_{{{\cal J}_c}}}\left( {{\mathbf{q}_{i,c}}^*,{\mathbf{p}_{i,c}}} \right)}}{{\partial {p_{i,c}^1}}} < 0.
\end{numcases}
Therefore, the maximum value of subscriber CAV group $\mathcal{J}_c$'s utility function can be derived by solving $\frac{{\partial {\mathcal{U}_{{{\cal J}_c}}}\left( {{\mathbf{q}_{i,c}}^*,{\mathbf{p}_{i,c}}} \right)}}{{\partial {p_{i,c}^1}}} = 0$ with KKT conditions, i.e.,
\begin{align}\label{eq:pic1}\resizebox{.894\hsize}{!}{$
{p_{i,c}^1}^* = {\frac{{\sqrt {\left({\xi_c^1}{\varepsilon _{i,\pi_c}^1}\right)^2 + {J_c^1}{\alpha _{i,c}}{f_c}{R_i}{\xi_c^1} {\varepsilon _{i,\pi_c}^1}{sc_{i,\pi_c}}}  - {\xi_c^1} {\varepsilon _{i,\pi_c}^1}}}{{{J_c^1}{\vartheta_c^1} }}}.$}
\end{align}
Here, we define $p_\theta^1 = \frac{2{\xi_c^1} {\varepsilon _{i,\pi_c}^1}{sc_{i,\pi_c}}}{{J_c^1{\vartheta_c^1} }}$ and consider two cases.

Case $1$: If $0<{p_{i,c}^1}^* < p_\theta^1$, we can observe that the utility function ${\mathcal{U}_{{\mathcal{J}_c}}}\left( {{{\bf{q}}_{i,c}}^*,{{\bf{p}}_{i,c}}} \right)$ increases when $p_{i,c}\in \left[0,{p_{i,c}^1}^*\right]$ while decreases when $p_{i,c}\in \left[{p_{i,c}^1}^*,p_{\max}\right]$. Therefore, ${p_{i,c}^1}^*$ is the optimal payment strategy, as shown in Eq. (\ref{eq:pic1}). In this case, by solving $0<{p_{i,c}^1}^* < p_\theta^1$, the constraint can be derived as below:
\begin{align}
J_c^1{\alpha _{i,c}}{f_c}{R_i} - 4\xi _c^1\varepsilon _{i,{\pi _c}}^1\left( {s{c_{i,{\pi _c}}} + 1} \right) < 0.
\end{align}

Case $2$: If ${p_{i,c}^1}^* \ge p_\theta^1$, we can derive that the utility function ${\mathcal{U}_{{\mathcal{J}_c}}}\left( {{{\bf{q}}_{i,c}}^*,{{\bf{p}}_{i,c}}} \right)$ increases when $p_{i,c}\in \left[0,p_\theta^1\right]$ while decreases when $p_{i,c}\in \left[p_\theta^1,p_{\max}\right]$. Hence, $p_\theta^1$ is the optimal payment strategy. In this case, by solving ${p_{i,c}^1}^* \ge p_\theta^1$, we have the following constraint as:
\begin{align}
J_c^1{\alpha _{i,c}}{f_c}{R_i} - 4\xi _c^1\varepsilon _{i,{\pi _c}}^1\left( {s{c_{i,{\pi _c}}} + 1} \right) \ge 0.
\end{align}
Theorem $2$ is proved.
\end{IEEEproof}

\section{{proof of theorem $3$}}\label{Appendix C}
\begin{IEEEproof}
According to the backward induction approach, the SE of the game $\mathbb{G}$ can be denoted as
\begin{align}
\left({\mathbf{p}_{i,c}^u}^*,{\mathbf{q}_{i,c}^u}^*\right) = \left({\mathbf{p}_{i,c}^u}^*,r\left({\mathbf{p}_{i,c}^u}^*\right)\right), \forall u \in \{1,2\},
\end{align}
where $r(.)$ is the optimal response function of publisher CAV $i$ defined in Eq. (\ref{eq: responsefunc}). Here, we only consider the case that $u=1$, and the other case that $u=2$ can be proved similarly. Next, we consider the following two cases.

Case $1$: $\Psi^1  \ge 0$. In this case, according to Theorem $1$ and Theorem $2$, we have ${p_{i,c}^1}^* = \frac{{2}{\Lambda^1}}{{{J_c^1}{\vartheta_c^1} }}$ and ${q_{i,c}^1}^* = 1$.

Case $2$: $\Psi^1  < 0$. In this case, from Theorem $1$ and Theorem $2$, we have ${p_{i,c}^1}^* = \frac{{\sqrt {\Upsilon^1}   - \Omega^1}}{{{J_c^1}{\vartheta_c^1} }}$ and
\begin{align}
{q_{i,c}^1}^* = r\left({p_{i,c}^1}^*\right) = \frac{{J_c^1{\vartheta_c^1} }}{{2{\xi_c^1} {\varepsilon _{i,\pi_c}^1}{sc_{i,\pi_c}}}}{p_{i,c}^1}^* = \frac{{\sqrt {\Upsilon^1} - \Omega^1}}{2\Lambda^1}.
\end{align}
Accordingly, the SE of the game $\mathbb{G}$ can be attained, as shown in Eq. (\ref{eq:stackSE}). Theorem $3$ is proved.
\end{IEEEproof}

\end{appendices}

\bibliographystyle{IEEETran}
\bibliography{SPAD_ref}
\begin{IEEEbiography}[{\includegraphics[width=1in,height=1.25in,clip,keepaspectratio]{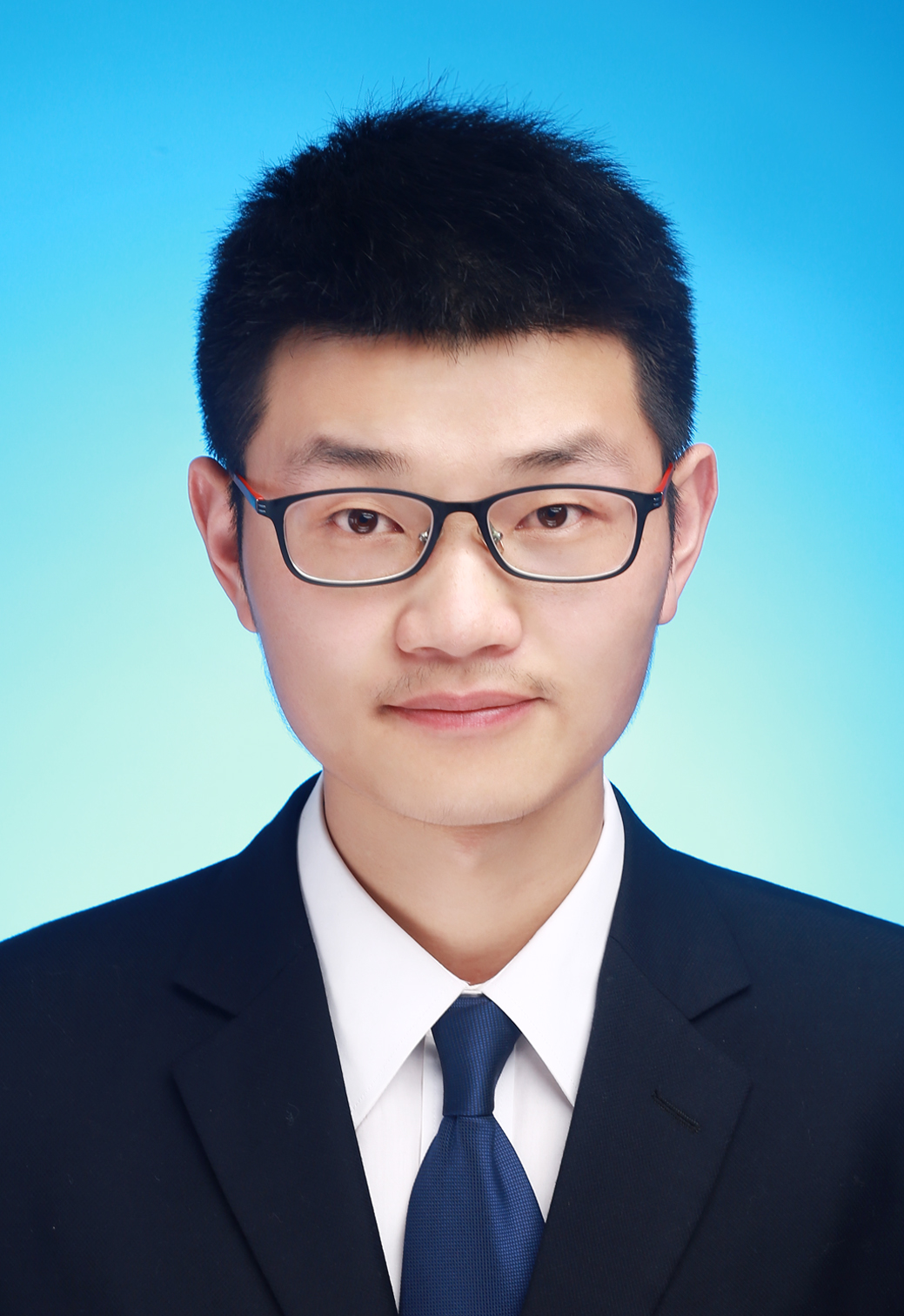}}]{Yuntao Wang}
received the Ph.D degree in Cyberspace Security from Xi'an Jiaotong University, Xi'an, China, in 2022, where he is currently an Assistant Professor with the School of Cyber Science and Engineering. His research interests include security and privacy in intelligent IoT, network games, and blockchain.
\end{IEEEbiography}\vspace{-1cm}

\begin{IEEEbiography}[{\includegraphics[width=1in,height=1.25in,clip,keepaspectratio]{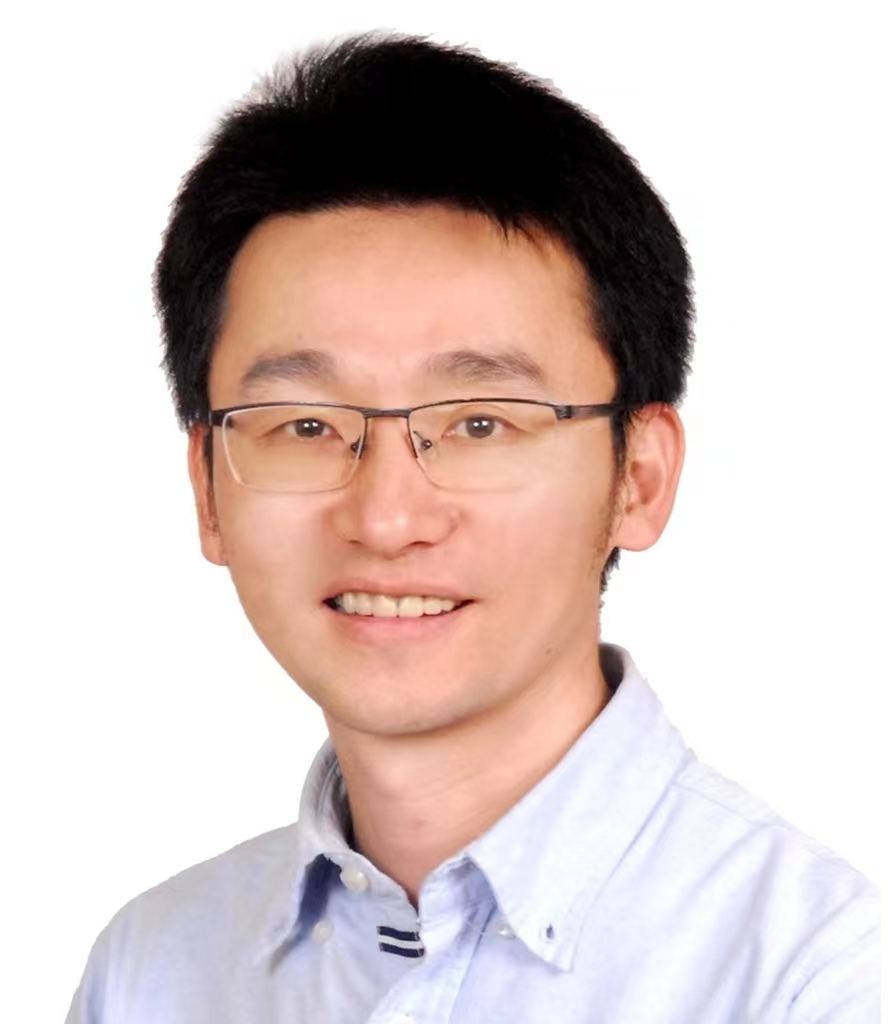}}]{Zhou Su}
has published technical papers, including top journals and top conferences, such as {\scshape IEEE JOURNAL ON SELECTED AREAS IN COMMUNICATIONS}, {\scshape IEEE TRANSACTIONS ON INFORMATION FORENSICS AND SECURITY}, {\scshape IEEE TRANSACTIONS ON DEPENDABLE AND SECURE COMPUTING}, {\scshape IEEE TRANSACTIONS ON MOBILE COMPUTING}, {\scshape IEEE/ACM TRANSACTIONS ON NETWORKING}, and {\scshape INFOCOM}. His research interests include multimedia communication, wireless communication, and network traffic. Dr. Su received the Best Paper Award of International Conference IEEE ICC2020, IEEE BigdataSE2019, and IEEE CyberSciTech2017. He is an Associate Editor of {\scshape IEEE INTERNET OF THINGS JOURNAL}, {\scshape IEEE OPEN JOURNAL OF COMPUTER SOCIETY}, and {\scshape IET Communications}.
\end{IEEEbiography}\vspace{-1cm}

\begin{IEEEbiography}[{\includegraphics[width=1in,height=1.25in,clip,keepaspectratio]{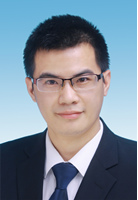}}]{Qichao Xu}
received the Ph.D degree from the school of Mechatronic Engineering and Automation, Shanghai University, Shanghai, China, in 2019. He is currently an Associate Professor with Shanghai university. His research interests are in trust and security, the general area of wireless network architecture, internet of things,  vehicular networks, and resource allocation. He has published more than 50 papers in some respected journals, e.g., IEEE TIFS, IEEE TDSC, IEEE TWC, IEEE TII, IEEE TVT, etc. He was receipt of the best paper awards from several international conferences including IEEE IWCMC2022, IEEE MSN2020, EAI MONAMI2020, IEEE Comsoc GCCTC2018, IEEE CyberSciTech 2017, and WiCon2016.
\end{IEEEbiography}

\begin{IEEEbiography}[{\includegraphics[width=1in,height=1.25in,clip,keepaspectratio]{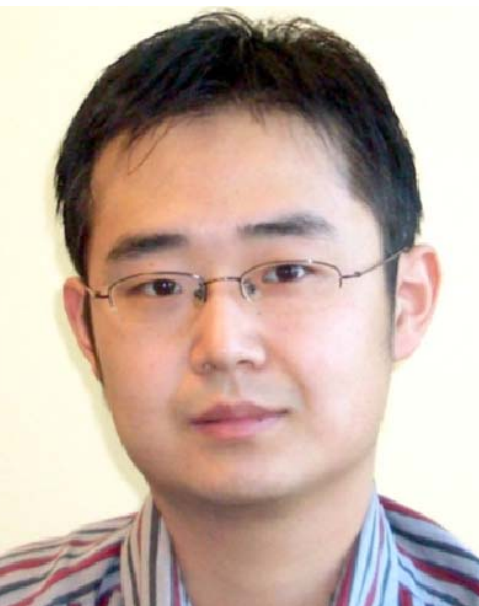}}]{Tom H. Luan}
received the Ph.D. degree from the University of Waterloo, Canada, in 2012. He is currently a Professor with Xi'an Jiaotong University, China. He has authored/coauthored more than 97 journal articles and 58 technical articles in conference proceedings. He awarded one U.S. patent. His research mainly focuses on content distribution and media streaming in vehicular ad hoc networks and peer-to-peer networking and the protocol design and performance evaluation of wireless cloud computing and edge computing.
\end{IEEEbiography}\vspace{-1cm}

\begin{IEEEbiography}[{\includegraphics[width=1in,height=1.25in,clip,keepaspectratio]{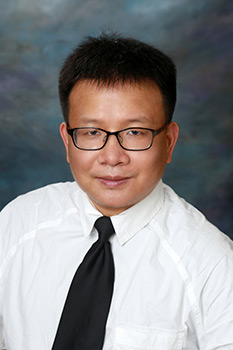}}]{Rongxing Lu}
(S'09-M'11-SM'15-F'21) is currently an Associate Professor at the Faculty of Computer Science (FCS), University of New Brunswick (UNB), Canada. Before that, he worked as an assistant professor at the School of Electrical and Electronic Engineering, Nanyang Technological University (NTU), Singapore from April 2013 to August 2016. He worked as a Postdoctoral Fellow at the University of Waterloo from May 2012 to April 2013. He was awarded the most prestigious ``Governor General's Gold Medal", when he received his PhD degree from the Department of Electrical \& Computer Engineering, University of Waterloo, Canada, in 2012; and won the 8th IEEE Communications Society (ComSoc) Asia Pacific (AP) Outstanding Young Researcher Award, in 2013. Dr. Lu is an IEEE Fellow. His research interests include applied cryptography, privacy enhancing technologies, and IoT-Big Data security and privacy. He has published extensively in his areas of expertise, and was the recipient of 8 best (student) paper awards from some reputable journals and conferences. Currently, Dr. Lu serves as the Chair of IEEE ComSoc CIS-TC (Communications and Information Security Technical Committee), and the founding Co-chair of IEEE TEMS Blockchain and Distributed Ledgers Technologies Technical Committee (BDLT-TC). Dr. Lu is the Winner of 2016-17 Excellence in Teaching Award, FCS, UNB.
\end{IEEEbiography}
\end{document}